\documentclass[namedreferences]{solarphysics}

\usepackage[hyperref,optionalrh]{spr-sola-addons}
\usepackage{graphicx}        

\usepackage{color}           

\newcommand{\aap}{    {\it Astron. Astrophys.}}

\chardef\us=`\_

\begin{document}

\begin{article}
\begin{opening}

\title{Adding further pieces to the synchronization puzzle: QBO, bimodality, and phase jumps}


\author[addressref={aff1},corref,email={F.Stefani@hzdr.de}]{\inits{F.}\fnm{F.}~\lnm{Stefani}}\sep
\author[addressref={aff1}]{\fnm{G.M.}~\lnm{Horstmann}}\sep
\author[addressref={aff1}]{\fnm{G.}~\lnm{Mamatsashvili}}\sep
\author[addressref={aff1}]{\fnm{T.}~\lnm{Weier}}\sep
\address[id=aff1]{Helmholtz-Zentrum Dresden -- Rossendorf, Bautzner Landstr. 400,
D-01328 Dresden, Germany}

\runningauthor{F. Stefani {\it et al.}}
\runningtitle{Adding further pieces to the synchronization puzzle: QBO, bimodality, and phase jumps}

\begin{abstract} This work builds on a recently developed self-consistent
synchronization model of the solar dynamo which attempts to explain
Rieger-type periods, the Schwabe/Hale cycle and the Suess-de Vries
and Gleissberg cycles in terms of resonances of various wave 
phenomena with gravitational forces exerted by the orbiting planets.
We start again from the basic concept that the spring tides of the three
pairs of the tidally dominant planets Venus, Earth and Jupiter 
excite magneto-Rossby waves at the solar tachocline. 
While the quadratic action of the sum of these three waves comprises 
the secondary beat period of 11.07 years, the main focus is 
now on the action of the even more pronounced
period of 1.723 years. Our dynamo model
provides oscillations with 
exactly that period, which is
also typical for the
quasi-biennial oscillation (QBO). Most remarkable
is its agreement with 
Ground Level Enhancement (GLE) events which preferentially
occur in the positive phase
of an oscillation with a period of 1.724 years.
While bimodality of the sunspot distribution is shown to be
a general feature of synchronization,
it becomes most strongly expressed under the influence of the QBO. 
This  may explain
the observation that the solar activity is relatively subdued 
when compared to that of other sun-like stars.
We also discuss anomalies of the solar cycle, and subsequent phase jumps by 180$^{\circ}$.
In this connection it is noted 
that the very 11.07-year beat period 
is rather sensitive to the time-averaging of the quadratic functional of the waves and 
prone to phase jumps of 90$^{\circ}$. 
On this basis, we propose an
alternative explanation of
the observed 5.5-year phase jumps
in algae-related data from the North Atlantic and Lake Holzmaar
that were hitherto 
attributed to optimal growth conditions.

\end{abstract}
\end{opening}
\section{Introduction}

For more than six decades, modeling of the solar magnetic field has mainly relied on the $\alpha-\Omega$-dynamo concept,
where $\Omega$ stands for the differential rotation which winds up toroidal from poloidal field, while the $\alpha$-effect replenishes the poloidal from the toroidal field, thereby 
closing the dynamo cycle. When adding to those 
mechanisms the induction effect of
the meridional circulation in the convection zone, 
contemporary dynamo models have become quite skillful in explaining both the butterfly 
diagram of sunspots and their 11-year intensity cycle \citep{Charbonneau2020}.
The recognition of that periodicity by Schwabe in 1843 \citep{Schwabe1844} 
is an astounding example of a late discovery, both in
the biographical and the historical dimension.
In light  of Galilei's and Scheiner's use of the telescope in the early 17$^{th}$ century to observe sunspots,
one could have expected a much earlier identification of the Schwabe cycle. 
Obviously, though, the paucity of sunspots
throughout the Maunder minimum (approx. 1645-1715) was not supportive in this respect. 

Soon after Schwabe's discovery the question of whether ``his'' cycle is somehow connected with the motion of the planets was brought up by \cite{Wolf1859}, and has re-appeared in the literature every now and then \citep{delarue1872,Bollinger1952,Jose1965,Takahashi1968,Wood1972,Okal1975,Condon1975,Charvatova1997,Landscheidt1999,DeJager2005,Charbonneau2022}. 
Yet, it was not
until recently that various authors \citep{Hung2007,Scafetta2012,Wilson2013,Okhlopkov2016} argued for a key role of the
11.07-year alignment period of the tidally dominant planets Venus, Earth and Jupiter in setting the Schwabe cycle. 

On the first glance, the focus on such 
high accuracy numbers may seem preposterous, given the broad
variability of the length of the Schwabe cycle,
which is subject not only to noise but also to 
systematic periodicities on the 
centennial time scale \citep{Richards2009,Chatzi2023}.
From this viewpoint, even a perfect agreement between 
two mean periods might still be pure coincidence.
However, quite another perspective opens up when taking into account
the possibility of clocking and synchronization. Asking 
``Is there a chronometer hidden deep in the Sun'', \cite{Dicke1978}
had introduced the ratio between the mean square of the
residua of the instants of
cycle maxima/minima from a long-term trend to that
of the difference between two subsequent residua.
Considering a time series with $N$ events, 	
Dicke's ratio is a viable measure for distinguishing between
random walk  and clocked processes, 
converging to $N/15$ 
for the former and to 0.5 for the latter.
For the clear identification of synchronized processes, the accurate determination of 
the mean-cycle length becomes indeed imperative, even if the individual cycles show  significant variability.
Only if synchronization can plausibly be assumed, one 
might ask then 
what physical mechanisms could
be responsible for it.

Our own contributions to this debate started in 2016 with the idea that the $\alpha$-effect in the
solar tachocline (thought to be related to the Tayler instability therein, see \cite{Weber2015}) could be synchronized by a presumed
11.07-year tidal forcing as exerted by the above-mentioned planets \citep{Stefani2016,Stefani2018}. The implementation of this concept  in a simple ordinary
differential equation (ODE) model of an $\alpha-\Omega$-dynamo suggested that any conventional solar dynamo, with a
``natural'' period not too far from 22.14 years, would be entrained to exactly this value if only the periodic part
of the $\alpha$-effect was strong enough. In a follow-up paper \citep{Stefani2019}, a 1D partial differential equation (PDE) model (in
co-latitude $\theta$ and time $t$) showed the same parametric-resonance type entrainment effect. Still later, 
this 1D model was enhanced by incorporating the 19.86-year period of the rosette-shaped motion of
the Sun around the barycenter of the solar system into a field-storage capacity term \citep{Stefani2020a}. The resulting double synchronization yielded 
a dominant 193-year beat period
(as anticipated previously by  \cite{Wilson2013} and \cite{Solheim2013}), together with some Gleissberg-type periods, followed by a transition into chaos
\citep{Stefani2021}. But only in 2024, the power spectrum resulting 
from this relatively simple model was found \citep{Stefani2024} to be in astonishing agreement with that
of climate-related sediment data \citep{Prasad2004} from Lake Lisan\footnote{Concerning the unavoidable question of where the relatively strong climatic impact  of solar activity stems from (which is beyond the scope of the present paper), see \cite{Gray2010} and \cite{Georgieva2023}. 
The finding of \cite{Veretenenko2023} that, in the period 
between 1873 and 2021,
cyclone tracks in the North Atlantic were shifted northward under a secular 
lowering of solar activity and southward under its enhancement, 
could also be important for explaining the variability of  
precipitation rates in the Lake Lisan area, as
already 
discussed by \cite{Prasad2004}). For the important role of the Suess-de Vries and Gleissberg cycles in solving the multivariate
attribution problem of solar activity and CO$_2$, see \cite{Stefani2021b}.}. When glancing, in Figure 9 of \cite{Stefani2024}, at the precise agreement between the two resulting Suess-de Vries peaks at 193 and 192 years, respectively, it is hard to sustain the belief in pure coincidence.

Despite those promising results of the
simple ODE and 1D-PDE models, they were received in the solar-dynamo community with mixed response. And understandably so, in view of three
plausible counter-arguments that were promptly brought forward by critics:
\begin{enumerate}
\item The presumed phase stability, i.e. the clocking of the Schwabe cycle, was put into serious question
\citep{Nataf2022}. In particular, the somewhat antiquated cycle data once corroborated by \cite{Schove1955} (which the analysis of \cite{Stefani2019} relied on), were
criticized as being ``finagled'' by 
his 9-per-century rule which would automatically generate a clocked
process with a period of 11.11 years. In a similar spirit, utilizing newer $^{14}$C data of \cite{Brehm2021}, \cite{Weisshaar2023} made a stark claim 
in favour of a non-clocked, random-walk-type nature of the solar cycle.
\item A criticism with a long tradition refers to the small magnitude of the tidal forces of the planets, which
amount to a tidal height (at the tachocline level) of less than 1 mm. It is indeed hard to conceptualize how this could have 
any relevant effect on the solar dynamo \citep{Callebaut2012,Nataf2022}.
\item The alleged 11.07-year period of the tidal forcing, a central tenet of our original synchronization concept, actually does not
show up in the tidal potential \citep{Nataf2022,Cionco2023}.
\end{enumerate}
These three pieces of criticism have recently prompted us to go significantly beyond our preliminary 
``toy models''. 
Here are the rebuttals we have meanwhile arrived at:
\begin{enumerate}
\item As for the 9-per-century rule, we showed that Schove's long-term data covering nearly two
millennia point indeed to a period of 11.07 years rather than of 11.11 years \citep{Stefani2020a}. More importantly, we
revealed in \cite{Stefani2023a} that the alleged disprove of phase-stability, as put forward by \cite{Weisshaar2023}, relies essentially on
the insertion of one additional solar cycle around 1845, whose existence might be inferred from the 
$^{14}$C data of \cite{Brehm2021} and \cite{Usoskin2021}, but for which there is no evidence from any direct observation. A similar erroneous cycle insertion
might have happened around 1650, i.e., amidst the Maunder minimum where cycles are notoriously hard to identify. If both of these additional cycles are removed, we end up again with a
clocked process back to the year 1140, vindicating the allegedly ``antiquated'' data of \cite{Schove1955,Schove1979,Schove1983,Schove1984}. Moreover, two independent sets of
algae-related data from a 1000-year interval in the early Holocene provided independent evidence for phase
stability \citep{Vos2004,Stefani2020b} with a period of 11.04 years (which is, in light  of the statistical uncertainty, hardly distinguishable
from 11.07 years)\footnote{Notably, these important algae-related data were up to now completely ignored
by all critics of synchronization.}. While we do not go so far as to claim unassailable evidence for clocking, we reject claims that conclusive proof exists for a non-clocked process.
\item Inspired by recent research on magneto-Rossby waves in the solar tachocline \citep{Zaqarashvili2010,Zaqarashvili2018,Dikpati2018,Dikpati2020,Dikpati2021a,Dikpati2021b,Raphaldini2019,Raphaldini2022,Raphaldini2023,Marquez2017}, we have asked
how these waves would react on external tidal triggers \citep{Horstmann2023}. Fortunately, the derived closed equation
for the wave's velocity component in meridional direction allowed for a quasi-analytical solution of this
problem. For the three periods of the two-planet spring tides, which are 118 days (Venus-Jupiter), 199
days (Earth-Jupiter) and 292 days (Venus-Earth),\footnote{For the sake of shortness, we initially stick to those approximate numbers like 118, 199 and 292 days; the precise
values will be provided in Section 4.} we arrived at wave velocities between 0.1 and
100 m/s, depending on the instantaneous value of the toroidal magnetic field and a poorly constrained damping parameter \citep{Stefani2024}. Such a resulting velocity 
scale (which is indeed comparable with other dynamo-relevant velocities) is actually not that surprising when translating the tidal height of $h \approx 1$\,mm 
into an  energetically
equivalent velocity of $(2 g_{\rm tacho} h)^{1/2}=1$\,m s$^{-1}$ 
(with the gravitational  acceleration at the tachocline 
level of $g_{\rm tacho}=500$\,m s$^{-2}$).
Yet, to make this tidal energy accessible,
it requires an appropriate ``resonance ground''
that is indeed provided by the (magneto-)Rossby waves.
Interestingly, their periods correspond to so-called Rieger-type
periodicities (originally identified as 154 days \citep{Rieger1984}), which can be found in various proxies of solar activity \citep{Knaack2005,Gurgenashvili2016,Gurgenashvili2021,Gachechiladze2019,Bilenko2020,Korsos2023}.
\item The problem related to the non-appearance of the very 11.07-yr signal in the tidal potential was resolved
by analyzing the collective effect of the three individual magneto-Rossby waves \citep{Stefani2024}. Figure 8 in this paper showed the
azimuthal average of their squared sum, which clearly comprises an 11.07-yr beat period. This belongs to the class of so-called {\it secondary beat periods}, which never appear in Fourier spectra, but frequently show up in nature. Perhaps the most prominent example originates from music theory, where secondary beats appear in slightly mistuned consonant intervals and chords \citep{Plomp1967,Nuno2024}. 
The application of this concept to our problem will be discussed in Appendix A.   
While a thorough computation of the 
$\alpha$-effect that {\it results} from the waves 
is still pending, it is reasonable to guess that it
will be in the order of 1/100...1/10 of the basic wave velocity. This, in turn, matches well to the estimated amplitude of
dm/s that would be {\it required} to entrain the entire solar dynamo \citep{Klevs2023}. The 2D solar dynamo model
utilized for this estimation, comprising (more or less) realistic profiles of the differential
rotation, the meridional circulation, and a conventional $\alpha$-effect in the convection zone, showed again the 
usual 
parametric resonance if an additional 11.07-year periodic $\alpha$-effect in the tachocline 
was added.
\end{enumerate}

In view of these new developments, we assert that the 
planetary synchronization theory of the Schwabe/Hale cycle stands now 
on rather solid ground. To a lesser extend, this applies also to the secondary synchronization mechanism, which is needed to
understand longer-term cycles. As already mentioned,
the additional period that comes into play at this point is the 19.86-year conjunction cycle of Jupiter and Saturn  that governs the rosette-shaped motion of the Sun around the barycenter of the solar system. Being often dismissed as irrelevant for the solar dynamo due to its ``free-fall'' character, recent work by \cite{Shirley2006,Shirley2017,Shirley2023} points to the excitation of an internal motion by virtue of spin-orbit coupling. Interestingly, the resulting motion, as shown in Figure 4  of \cite{Shirley2017}, exhibits the same $m = 1$ azimuthal dependence as is also typical for precession-driven flows \citep{Giesecke2024}. In view of the structural 
similarity of the spin-orbit coupling term (the ``CTA''-term in the parlance of \cite{Shirley2017}) with the Poincar\'{e} force in precession, 
this is hardly surprising. That said, we admit that a quantitative determination of the flow that is directly driven by spin-orbit coupling, and any axi-symmetric secondary flow emerging from it, is still elusive. 

In a first attempt to make this problem somehow tractable, we had incorporated the 19.86-year forcing into a model parameter that is related to the field-storage capacity in the tachocline. The rationale behind this choice is the 
high sensitivity of the adiabaticity in the tachocline 
with respect to external perturbations \citep{Ferrizmas1994,Abreu2012}. 
That way, we obtained 
the Suess-de Vries cycle as a beat period
of $22.14 \times 19.86/(22.14-19.86)=193$ years, as mentioned above\footnote{While the actual period
of the Suess-de Vries cycle is a matter of debate, and
often indicated as 208 years, our choice of 193 years 
is encouraged by the sharp 192-year 
peak obtained from the very long (8500 years) sediment data 
from Lake Lisan \citep{Prasad2004}. It gains further 
plausibility from similar results of 
\cite{Richards2009} (188 years) and \cite{Ludecke2015} (between 186 and 202 years)}.

\begin{figure}[t]
  \centering
  \includegraphics[width=0.99\textwidth]{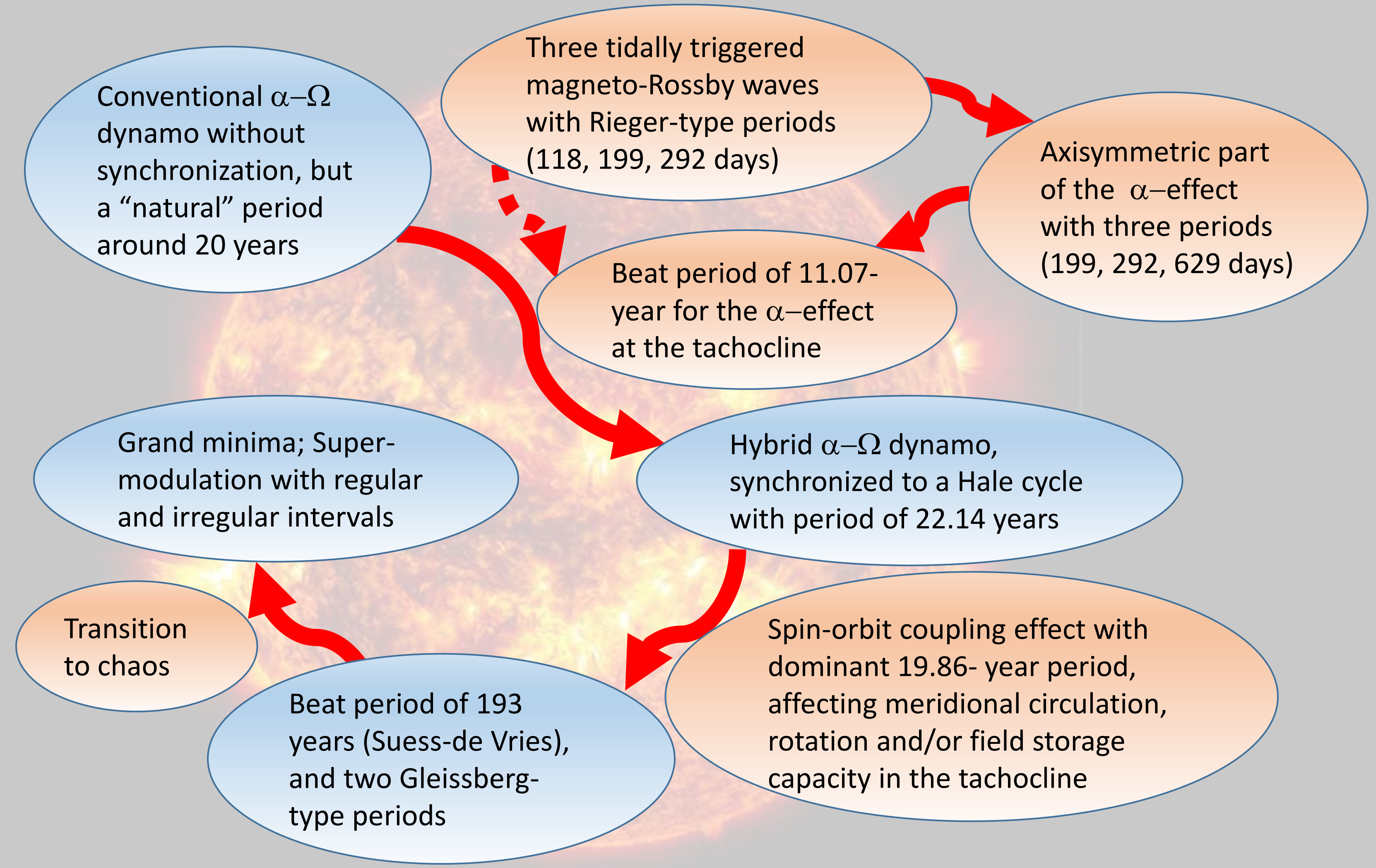}
  \caption{Scheme illustrating our present understanding of the double-synchronized solar dynamo. Light blue areas denote the temporal features of the dynamo, while pink areas denote the underlying physical mechanisms. A conventional solar dynamo is entrained by a 11.07-year periodic $\alpha$-effect in the tachocline that, in turn, emerges as the beat between three tidally excited magneto-Rossby waves. A secondary beat period of 193 years results then from the 22.14-year Hale cycle and the 19.86-year barycentric motion. Finally, supermodulation between regular and irregular 
  intervals \citep{Weiss2016} emerges after transition to chaos. The main focus (upper right part) of this paper lies now on the beat period of 629 days that will be shown to trigger a QBO.}
  \label{fig:3}
\end{figure}

Our present understanding of the double-synchronized solar dynamo is summarized in Figure 1. Besides the remarkable agreement of the resulting dominant periods with those inferred from climate data (see Figure 9 in \cite{Stefani2024}), another important feature is its high degree of self-consistency. Indeed, the sharp 193-year beat period {\it would not even emerge} if the two
underlying periods, 22.14 and 19.86 years, {\it were not phase-stable} in the first place. Or, to put it otherwise: the
hypothesis of a primarily clocked 11.07-year Schwabe cycle is strongly supported by the sharpness of the
secondary 193-year Suess-de Vries cycle. An analogous argument applies to the presumed clocking of
the 118, 199, and 292-day magneto-Rossby waves 
determining the phase stability of their 11.07-year secondary beat period. While we are open to discuss all sorts of reasonable arguments against this scheme, we are not aware of any other solar dynamo model that provides all the relevant periodicities in such a closed and self-consistent manner.

With this background, the main focus of this paper lies now on the upper right part of Figure 1. While the 
primary period of 1.723 years (629 days)
was shortly mentioned in \cite{Stefani2024}, our attention was (too?) quickly shifted  to the secondary beat period of 11.07 years (i.e., along the dashed red arrow in Figure 1).

Here we step back again and consider the axi-symmetric part of the quadratic action of the three waves in greater detail. We will show that this 
action comprises only three periods, namely 199 days, 292 days, and $629 \approx 199 \times 292/(292-199)$ days (1.723 years). While the
118-days period vanishes in the quadratic functional, the 
11.07-year period survives as a secondary beat period as will be shown in the Appendix.

The experienced solar scientist has certainly noticed that the 1.723-year period
is suspiciously close to the typical value of the
so-called quasi-biennial oscillation (QBO). Since 
its discovery in solar holes by \cite{McIntosh1992}, 
and in cosmic rays and open magnetic flux by
\cite{Valdez1996} and \cite{Rouillard2004}, who 
found a dominant period of 1.68 years, this oscillation
has acquired a lot of interest in recent decades 
\citep{Bazilevskaya2014,Kiss2018}. Similar periodicities, although
closer to 1.3 or 1.5 years, had also been observed in sub-surface
flows \citep{Howe2000,Inceoglu2021}. 
Most interesting, since very close to our value,  is the 1.73-year period
that was observed by \cite{Velasco2018} in their analysis
of Ground Level Enhancement (GLE) events.
In a numerical effort to understand 
their physical nature, \cite{Inceoglu2019}
found QBO-like intermittent oscillations 
and a certain hemispheric decoupling
when employing a turbulent $\alpha$-effect
just above the tachocline (while a non-local Babcock-Leighton-type 
dynamo generally failed to reproduce those features).
This is an encouraging basis for the present work where, however, 
the fluctuations of $\alpha$ will be linked to the
action of tidal forces on magneto-Rossby waves. Another encouragement stems 
from the work of 
\cite{Edmonds2021} who discussed the parallelity
of the QBO with various planetary alignments.

One of the open questions in this context is whether the Gnevychev gap \citep{Gnevyshev1967,Gnevyshev1977}, i.e. the emergence of two peaks of maximum activity separated by 2-3 year (a feature present in large sunspots, major flares, coronal
green-line emission and geomagnetic activity), is just a particularly dominant feature  
of the QBO, or perhaps a separate phenomenon. Closely related to this question is the interpretation 
of a bimodal behaviour of the solar activity, as discussed by \cite{Nagovitsyn2016} and \cite{Nagovitsyn2017}. In view of these uncertainties, we will - before entering the very QBO-topic - assess whether the Gnevyshev gap and the bimodal distribution appear already in the previous, simplified version of our
synchronization model, which starts 
immediately from an assumed 11.07-year periodicity of $\alpha$ in the tachocline.

The main part of this paper is then 
dedicated to the emergence of the QBO and the consequences following from it.
Our extended dynamo model will be shown to result in
a strong peak at 1.723 years (629.29 days), together with two neighboring peaks at 1.598 and 1.868 years which arise as side-bands from the modulation with the 22.14-year Hale cycle. 

As for the 11.07-year Schwabe cycle, it still comes about as the 
beat between the
199, 292 and 629-day periods, as  is easily 
visible in the envelope of their sum. 
Moreover, the bimodal distribution will arise
as a very typical feature of the QBO, as
the field strength - in particular around the maxima of the 
cycle - vaccillates between a high and a low state.

This way, the QBO leads naturally 
to a certain ``sedation'' of the solar dynamo which tends to spend less time at the highest field strength 
than it would do in case of a single-frequency oscillation. It seems worthwhile to consider this as a 
possible explanation of the recent observation that the Sun is relatively quiet, when compared to the activity of other sun-like stars \citep{Schaefer2000,Reinhold2020}. 
We may arrive here at a simple justification for the 
provocative hypothesis ``that the ``Sun is not `Sun-like' ''
\citep{Cliver2022}, possibly with grave 
ramifications for the habitability of the Earth.

Apart from those features, our new model version, when appropriately
enhanced by a parametrized effect of spin-orbit coupling, 
will show a very similar long-term behaviour as the previous version, 
including the dominant Suess-de Vries peak at  193-years.

At last, we will discuss various types of phase jumps of the solar cycle. 
First we will numerically illustrate the emergence of 
certain anomalies after which the solar cycle reappears with phase 
shifts of either 0$^{\circ}$ or 180$^{\circ}$. 
Yet, a surprise is also 
on offer here.
While the squared action of the three waves was averaged only 
over the azimuth, any physically relevant functional  ($\alpha$, zonal flow) will certainly also 
include some average over time. Admittedly, we have not yet a really detailed model for that, so 
we simply tested different time averages. Somewhat 
surprisingly, depending on the width of the averaging window, we observe a shift of the maximum 
of the 11.07-year envelope by half of that period. Motivated by similar phase-jumps as observed 
in solar-cycle related algae data by \cite{Vos2004},  we 
carry out simulations  with a periodically changing phase-shift of $\alpha$ which indeed result 
in corresponding phase shifts of the dynamo.  

Before discussing those issues in detail, in the next 
section we present our numerical model.

\section{Numerical model}

In the most recent simulations \citep{Klevs2023,Stefani2024}
we had employed a 2D $\alpha-\Omega$-dynamo code that is
quite similar to those used in the 
classical solar mean-field dynamo benchmark of
\cite{Jouve2008}.
We used rather realistic values and 
profiles of differential rotation and meridional circulation, 
and assumed a  value for $\alpha$
in the convection zone of appr. 1.3\,m/s. 
For the radial profile of the magnetic diffusivity
$\eta$ we assumed a 100-fold enhancement between the 
``quite'' tachocline  and the turbulent convection zone.
Among the wide range of $\eta$-values
found in the literature,
we have chosen $\eta_t= 2.13 \times 10^{11}$cm$^2$ s$^{-1}$, to keep the period of the 
{\it{non-synchronized}}
dynamo not too far from 22 years.

Here, however, we ``regress'' to the simple 1D $\alpha-\Omega$-dynamo 
code as used in \cite{Stefani2019} and \cite{Stefani2021},
which was inspired by a similar scheme 
of \cite{Jennings1991}.
An obvious drawback of this code is that it is
neither capable of accounting for radial
dependencies of its ingredients ($\alpha$, $\eta$, etc.)
nor of including the meridional circulation
which is known to play a key role in setting the dynamo cycle 
\citep{Charbonneau2000}. On the positive side, this simple code
allows for extensive tests of various parameter combinations 
as well as for very long simulations, and hence accurate determinations 
of spectra. Those were indeed important to 
reveal the remarkable agreement of the main peaks with 
those found in climate-related data from lake Lisan \citep{Prasad2004}.

In order to make the paper self-contained we 
reiterate here the essential features of this 1D code.
Basically we solve the same system of PDEs as in 
\cite{Stefani2019} and \cite{Stefani2021}, with the solar co-latitude 
$\theta$ representing the only spatial coordinate.

As usual, the axi-symmetric part of the solar
magnetic field is split into a
poloidal component ${\bf{B}}_P=\nabla \times (A {\bf{e}}_{\phi})$
and a toroidal component ${\bf{B}}_T=B {\bf{e}}_{\phi}$. Then 
the 1D PDE system reads
\begin{eqnarray} 
  \frac{{\partial} B(\theta,t)}{{\partial} t} &=& \omega(\theta,t) \frac{\partial A(\theta,t)}{\partial \theta} 
  + \frac{\partial^2 B(\theta,t)}{\partial \theta^2} -\kappa(t) B^3(\theta,t) \\
    \frac{{\partial} A(\theta,t) }{{\partial} t} &=& \alpha(\theta,t) B(\theta,t) 
    + \frac{\partial^2 A(\theta,t)}{\partial \theta^2} , 
   \end{eqnarray}
where $A(\theta,t)$ represents the vector potential of the 
poloidal field at co-latitude $\theta$ (running between 0 and $\pi$) 
and time $t$, and $B(\theta,t)$ represents the corresponding toroidal field. 
The dynamo is conventionally driven by the helical turbulence parameter $\alpha$ and
the radial derivative $\omega=\sin\theta d (\Omega r)/dr$
of the rotation profile, whereby
$\alpha$ and $\omega$ 
stand for the non-dimensionalized
versions of their dimensional counterparts $\alpha_{\rm dim}$ and 
$\omega_{\rm dim}$, related via $\alpha=\alpha_{\rm dim} R/\eta$  and
$\omega=\omega_{\rm dim} R^2/\eta$. In the following we will assume $R=5\times 10^8$\,m as the relevant radius of the
considered dynamo region and a higher value of
$\eta=7.16 \times 10^{11}$cm$^2$ s$^{-1}$ for the magnetic diffusivity.
Accordingly, the time is non-dimensionalized by
the diffusion time, i.e. $t=t_{\rm dim} \eta/R^2$, leading to  
110.7 years in our case.

This PDE system is solved by a finite-difference solver using the Adams-Bashforth method.
The initial conditions are chosen as
$A(\theta,0)=0$ and 
$B(\theta,0)=s \sin\theta+ u \sin2 \theta$,  
with the pre-factors $s=0.001$ and $u=1.0$ 
denoting symmetric and asymmetric 
components of the toroidal field. 
The boundary conditions at the North and South pole 
are $A(0,t)=A(\pi,t)=B(0,t)=B(\pi,t)=0$.

For the $\omega$-effect, we utilize the typical solar $\theta$-dependence
as given by \cite{Charbonneau2020},
\begin{eqnarray}
\omega(\theta)&=&\omega_0 (1-0.939-0.136 \cos^2\theta-0.1457 \cos^4\theta )\sin\theta
\end{eqnarray}
with $\omega_0=10000$.

The $\alpha$-effect consists of two main parts, $\alpha=\alpha^c+\alpha^p$, where 
\begin{eqnarray}
\alpha^c(\theta,t)&=&\alpha^c_0(1+\xi(t))  
 \frac{1}{{1+q^c_{\alpha} B^2(\theta,t)}} \sin 2 \theta
\end{eqnarray}
is the ``classical'' part with a constant $\alpha^c_0$ and a noise term $\xi(t)$, defined by the correlator $\langle \xi(t) \xi(t+t_1) \rangle = D^2 (1-|t_1|/t_{\rm corr})
\Theta(1-|t_1|/t_{\rm corr})$. 
Note that, instead of the dynamical quenching 
equations that are
based on the buildup of magnetic helicity 
\citep{Field2002},
we employ here a rather simple, algebraic 
quenching of $\alpha$ which is considered sufficient for 
our purposes. Note further that with the values of $R$ 
and $\eta$ given above, we can always regain the
dimensional $\alpha$ from its dimensionless counterpart 
via $\alpha_{\rm dim}=0.1432 \times \alpha$\,m s$^{-1}$.

The second contribution 
$\alpha^p$ is a periodic function of time,
\begin{eqnarray}
\alpha^p(\theta,t)&=&\alpha^p_0 P[B(\theta),t]
{\rm{sgn}}(90^{\circ}-\theta) \tanh^2\left(  \frac{\theta/180^{\circ}-0.5}{0.2}  \right)
\; \mbox{for $55^{\circ}<\theta<125^{\circ}$} \nonumber \\
&=&0 \; \mbox{elsewhere} \; ,
\end{eqnarray}
where the term $P[B(\theta),t]$  incorporates the resonance-like reaction of $\alpha$ on the time-dependent tidal forcing.
While in previous work \citep{Stefani2019,Stefani2021}
this term was always assumed to have the
structure  $\sin(2 \pi t/t_{11.07}) B^2/(1+q^p_{\alpha} B^4)$
(with $t_{11.07}=11.07 \times \eta/R^2$  denoting the dimensionless
counterpart of the 11.07-year tidal forcing period) 
in the following also different versions will be tested.
Formally, the relation
$\alpha_{\rm dim}=0.1432 \times \alpha$\,m s$^{-1}$
applies also to this periodic part, although 
its interpretation is 
more subtle. This has to do with the fact
that in our 1D model we have to merge two 
different $\alpha$-contributions 
that actually work in regions with strongly different 
diffusivities. When assuming an 100-fold diffusivity 
contrast between the tachocline and the 
convection zone
(as, e.g., in the 2D model of \cite{Klevs2023}), and
partly 
compensating this by the smaller thickness of the tachocline, we may guess 
that the ``real'' physical $\alpha^p$ is 
approximately 10 times smaller than 
$\alpha_{\rm dim}$.

Motivated by ideas of \cite{Jones1983} and \cite{Jennings1991}, the  term $\kappa(t) B^3(\theta,t)$ in Equation (1)
had been employed in \cite{Stefani2021} to account 
for field losses owing to magnetic buoyancy, on the 
assumption that the escape velocity is 
proportional to $B^2$.
While we admit that the underlying concept of spin-orbit coupling 
of the angular momentum of the Sun around the barycenter
into some dynamo 
relevant parameter remains an open question (for ideas, besides those of \cite{Shirley2017,Shirley2023}, 
see \cite{Zaqa1997,Jucket2000,Palus2000,Javaraiah2003,Wilson2013,Sharp2013}),
we assume in the following the time-dependence of the parameter
$\kappa(t)$  to be proportional to that of the angular momentum. 
Since $\kappa(t)$  is related to 
the very sensitive adiabaticity in the tachocline,
which could be easily influenced by slight changes in the
internal rotation profile
\citep{Ferrizmas1994,Abreu2012}, its modification by some sort
of spin-orbit coupling seems rather plausible.

For the Sun's orbital angular momentum
we utilize the same data as previously, which had been computed from 
the DE431 ephemerides  \citep{Folkner2014} 
for a 30000-year interval.
This function is dominated by the 19.86-years synodes of Jupiter and Saturn,
to which further contributions, mainly from Uranus and 
Neptune, are added. 
Again, we will use the normalized version $m(t)$ of this 
angular momentum curve 
for parametrizing the time-variation of $\kappa(t)$
according to 
\begin{eqnarray}
\kappa(t)&=&\kappa_0+\kappa_1 m(t) \;.
\end{eqnarray}

\section{Bimodality in the simple synchronization model}

Bimodality had been shown to be an intrinsic feature of 
solar activity \citep{Nagovitsyn2016,Nagovitsyn2017}. A possible explanation of its occurrence was offered by \cite{Georgieva2011},
who distinguished two ways with different time-scales 
(diffusion and meridional circulation) on which poloidal flux can be transported towards the tachocline. Bimodality is often discussed in connection with the Gnevyshev gap, i.e. the subsequent 
emergence of two maxima with typically different field strengths. Yet, whether this Gnevyshev gap is just the most distinctive expression of the more general QBO, is a matter of debate. As a sort of reference 
model, to which the simulations including the QBO can later be compared, in the following we restrict ourselves to the previous model version comprising only the single 11.07-year periodicity in  $\alpha^p$.

We employ the numerical model discussed above, wherein we 
use the conventional resonance term
\begin{eqnarray}
 P[B(\theta),t]&=&\sin(2 \pi t/t_{11.07}) B^2(\theta)/(1+q^p_{\alpha} B^4(\theta)), 
\end{eqnarray}
together with the specific
parameters $\omega_0=10000$, $\alpha^c_0=15$,
$q^p_{\alpha}=0.2$, $q^c_{\alpha}=0.8$, $D=0.05$.
The time variation of the field-storage parameter 
is taken as $\kappa(t)=0.6+0.2 m(t)$. These values were chosen to make close contact with the model discussed in \cite{Stefani2021} and \cite{Stefani2024}.

\begin{figure}[t]
  \centering
  \includegraphics[width=0.99\textwidth]{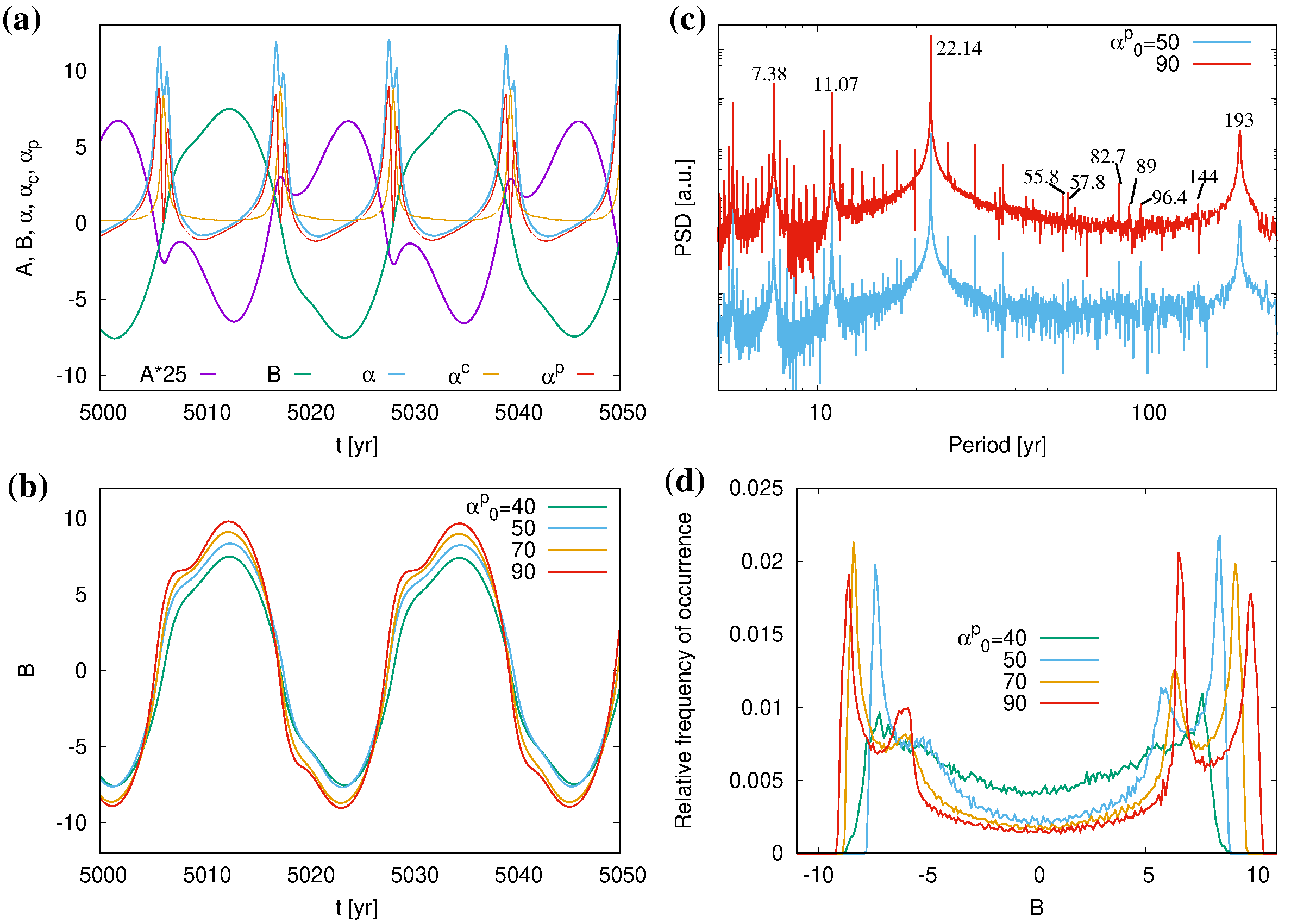}
  \caption{Emergence of bimodality in the simple synchronization model. (a) 
  Cutout of the time evolution of $A$, $B$, $\alpha$, and its two components $\alpha^c$ and $\alpha^p$, all measured at the co-latitude 
  $\theta=72^{\circ}$, when setting $\alpha^p_0=40$. Note the appearance of a strong and a weak ``hump'' 
  in $A$ and $B$, respectively. (b) Time evolution 
  of $B(\theta=72^{\circ})$ for four values $\alpha^p_0$ of the periodic forcing. (c) Spectra of $B$, for two values of $\alpha^p_0$ from (b), with the typical 
  dominant periods indicated. This type of spectrum had been shown in Figure 9 of \cite{Stefani2024} to remarkably agree with climate data from Lake Lisan.
  (d) Histograms of $B$ for four values $\alpha_0^p$ from (b), with the increasing second 
  peak. Note also the growing asymmetry of the histogram for larger $\alpha^p_0$.}
\end{figure}

What is varied then is the strength $\alpha^p_0$ of the periodic 
$\alpha$-term. 
For  $\alpha^p_0=40$, Figure 2(a) shows a 50-year segment of the time dependence of the two field 
components
$A$ and $B$, as well as of $\alpha$ and its two parts $\alpha^c$ and $\alpha^p$, all measured at 
co-latitude $\theta=72^{\circ}$.
For this value $\alpha^p_0=40$, the dynamo is already synchronized to 22.14 years.
While $\alpha^c$ shows a sharp peak for $B=0$ (since it is strongly quenched for non-zero $B$), 
$\alpha^p$ shows typical peaks shortly before and after
the transition point $B=0$, which 
reflects the 
resonance condition for a certain finite value of $B$. Obviously, this behaviour leads
to a noticeable ``hump'' of $A$, and a less pronounced, but still visible, 
flattening of $B$.
When increasing $\alpha^p_0$ from 40 to 90
(see Figure 2b), this feature becomes more pronounced, so that ultimately one might even 
speculate about the appearance of a 
Gnevyshev gap. When we compute the histogram of the value of $B$
over the computed 30000 years, we obtain Figure 2(d). 
The emergence of a double peak, the lower one at the $B$-values of the ``hump'', is clearly expressed,  indicating the bimodality in the probability distribution of $B$-field in this figure.
What is also seen is a certain symmetry breaking between negative and positive values, which regularly 
occurs in our model.

In Figure 2(c) we  add, for two values of $\alpha^p_0$, the resulting spectra of $B(t)$
with its dominant Hale and Suess-de Vries peaks at 22.14 and 193 years, jointly with some
Gleissberg-type peaks around 90-years and slightly below 60 years.
Basically, these are the same peaks as already shown in Figure 9 of \cite{Stefani2024}.

What we have learned so far is that even the simple model which produces Schwabe, Hale, Suess-de 
Vries and Gleissberg-type cycles
has a certain tendency to develop a second hump
and, thereby, a bimodal field distribution.
In the following we will see how this feature 
becomes even more pronounced when
the QBO is additionally taken into account.

\section{QBO, bimodality, and subdued activity}

We turn now to the central topic of this paper, the emergence of the QBO. Following \cite{Stefani2024}, we start
from the sum of the three tidally-excited magneto-Rossby waves
waves with periods 118 days, 199 days and 292 days,
\begin{eqnarray}
s(t)&=& \cos\left( 2\pi  \cdot \frac{t-t_{\rm VJ}}{0.5 \cdot P_{\rm VJ}}\right) +\cos\left( 2\pi  \cdot \frac{t-t_{\rm EJ}}{0.5 \cdot P_{\rm EJ}} \right)+\cos\left( 2\pi  \cdot \frac{t-t_{\rm VE}}{0.5 \cdot P_{\rm VE}} \right)
\end{eqnarray}
Herein,  we use the accurate two-planet synodic periods $P_{\rm VJ}=0.64884$\,years, 
$P_{\rm EJ}=1.09207$\,years,
$P_{\rm VE}=1.59876$\,years, and the 
epochs of the corresponding conjunctions
$t_{\rm VJ}=2002.34$,
$t_{\rm EJ}=2003.09$, and
$t_{\rm VE}=2002.83$
that were adopted from \cite{Scafetta2022}. For a 50-year and a 10-year interval,
this sum is shown in Figures 3a and 3f, respectively.
Anticipating that any dynamo-relevant quantity, be it
$\alpha$ or the $\Lambda$-effect \citep{Kitchatinov2005} 
driving zonal flows \citep{Tilgner2007,Morize2010} 
or meridional circulation \citep{Inceoglu2019}, will be a quadratic functional of those waves, 
for the sake of simplicity we consider just the square of the sum of the waves 
(Figures 3b and 3g). Assuming further that the most relevant dynamo contribution will likely be 
the axisymmetric part of this square, we average it over the azimuth, arriving at Figures 3c 
and 3h.  Evidently, this azimuthal average contains now only the old periods 199 and 292 days 
and the new one with $629 \approx 199 \times 292/(292-199)$ days, while the original 118-day period has disappeared\footnote{Strictly speaking, this only suggests that the 118-day period will not be very important for the global solar dynamo, while it could still appear in solar activity. Indeed, as shown in \cite{Gurgenashvili2021}, a signal at around 120 days is often visible in various solar activity indicators. As argued by \cite{Dikpati2021b}, toroidal magnetic flux is most likely to escape the 
tachocline in those places where the Rossby wave reaches its peak amplitude.}.

\begin{figure}[t]
  \centering
  \includegraphics[width=0.99\textwidth]{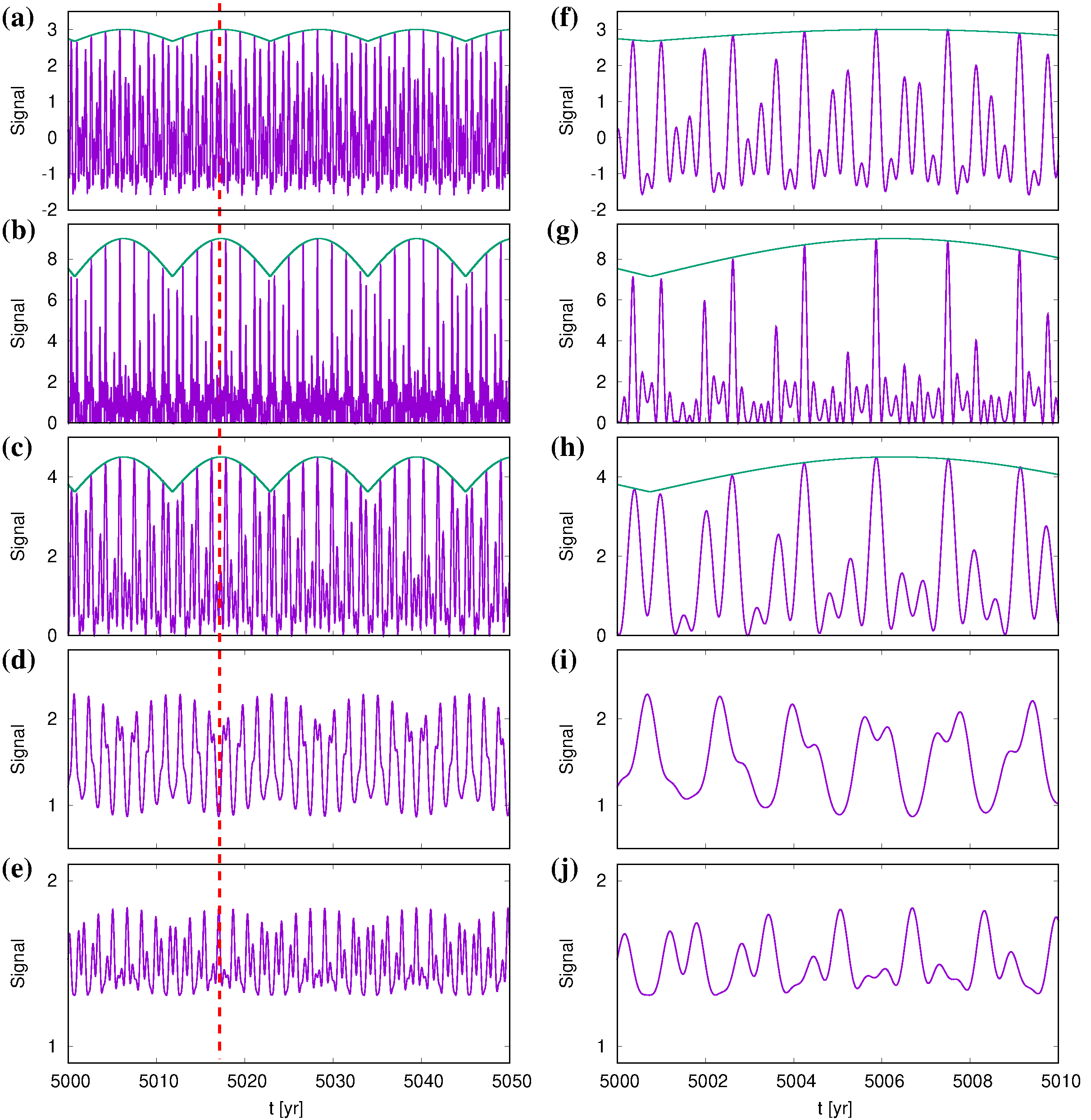}
  \caption{(a,f) The sum of the three waves with periods 118, 199, and 292 days for two 
  different time segments, according to Equation (8).
  (b,g) The square of this sum. (c,h) The azimuthal average over the square from (b,g).
  (d,i) Centered moving average of (c,h) with window 1.1 year. (e,j) Centered moving 
  average of (c,h) with window 2.1 year.
  The analytical expressions for the corresponding envelopes with period 11.07 years 
  (green curves) are derived in the Appendix A.}
\end{figure}

Up to this point, all functions were characterized by the well-known 11.07-year beat period 
which had played a central role in our synchronization model. The maximum of the envelope 
of all signals (derived in the Appendix) was consistently at the same position. 
This changes, however,  when we consider, in addition to the azimuthal average, also a time-average. The motivation for that lies in mean-field dynamo theory where the $\alpha$-tensor is calculated as an integral over space and time of a certain correlator of the velocity,
cp. Equation (5.51) in \cite{Krause1980}. While not going here into a detailed derivation of 
the $\alpha$-effect resulting from Rossby waves (see \cite{Avalos2009}) it is interesting to observe, 
from the difference between Figures 3d and 3e, that the maximum of the envelope shifts 
by 5.5 years when the widths of the moving average window is changing from 1.1 to 2.1 years. 
Before coming back to this effect later, we will first rely on the signal shown in 
Figure 3d that, after subtraction of the
mean, will be called $h(t)$ as a proxy for {\it{helicity}}.

In all following analyses, we will use a modified resonance term of the form 
\begin{eqnarray}
 P[B(\theta),t]&=&h(t) B^8(\theta)/(1+q^p_{\alpha} B^{16}(\theta)) \;.
\end{eqnarray}
The motivation for using  such high powers of $B$ is that the 11.07-year envelope seen in 
Figure 3 represents only a relatively minor variation when compared to the more dominant QBO. 
After having assessed
various forms of the resonance term we learned that the synchronization 
to 11.07 years requires a certain ``accentuation'' of this minor difference, which we 
accomplish by using a resonance term that is much steeper and more localized at higher 
values of $B$ than that of Equation (7). While this argument might seem a bit contrived, 
there is some physical rationale behind it in light of the much stronger wave excitation at 
relative high $B$-values (cp. Figures 2-4 in \cite{Stefani2024}). In this respect, one 
could also imagine to apply the $B$-dependent resonance condition to every term in the sum 
of the three waves separately. For the moment, however,  we stick to the simpler Equation (9).

\begin{figure}[t]
  \centering
  \includegraphics[width=0.99\textwidth]{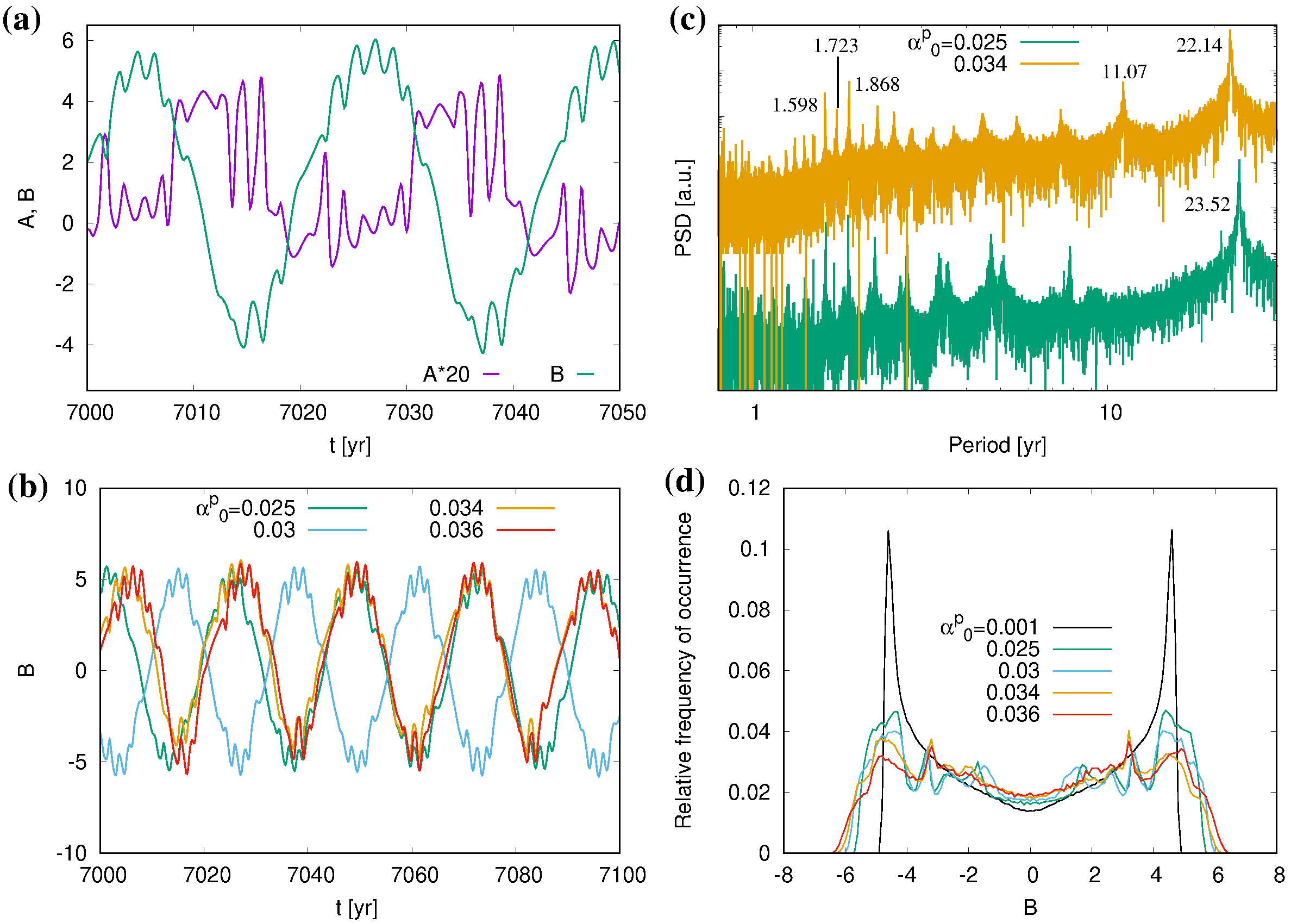}
  \caption{Emergence of QBO and bimodality in the synchronization model according to Equation (9). (a) 
  Cutout of the time evolution of $A$ and $B$, both measured at the co-latitude 
  $\theta=72^{\circ}$. Note the appearance of a 
  clear QBO in $B$ on top of the extrema of the main, longer period  oscillations. (b) Dependence of the time evolution 
  of $B(\theta=72^{\circ})$ on the strength 
  $\alpha^p_0$ of the periodic forcing. 
  For $\alpha^p_0=0.034$ and 0.036 the dynamo is synchronized to 
  22.14 years. For that reason the two curves 
  are widely parallel, while those for $\alpha^p_0=0.025$
  and 0.03 diverge in time.
  (c) Spectra of $B$ for two of the $\alpha_0^p$-values from (b), with the typical 
  periods indicated. Evidently we obtain a QBO peak at 
  1.723 years, together with two side bands emerging from the modulation by the 22.14-year Hale cycle.
  (c) Histograms of $B$ for four values of $\alpha_0^p$ from (b). 
  Note the remarkable flattening of those (colored) curves when compared with the (black) one for very weak periodic forcing.}
\end{figure}

For the new parameter choice $\omega_0=10000$, $\alpha^c_0=15$,
$q^p_{\alpha}=10^{-8}$, $q^c_{\alpha}=0.8$, $D=0.05$, $\kappa(t)=\kappa_0=0.6$, Figure 4 shows the resulting field evolution and the emerging spectra and histograms when increasing the periodic $\alpha$-term.
What we see first (in Figure 4a and 4b)  is a clear signature of the QBO with the emergence of two or even more local peaks around the cycle extrema.
Figure 4d shows the histograms of the $B$-field for the four values of $\alpha_0^p$ from (b). What is interesting here is the remarkable flattening of those (colored) curves when compared with the (black) one obtained for very weak periodic forcing.
 Figure 4c shows the spectra of $B$ for two values of $\alpha_0^p$ from (b), with the typical periods indicated. Evidently we obtain now a clear QBO peak at 
  1.723 years, together with its two side bands at 1.598 and 1.868 years that obviously 
  emerge
  from the modulation with the 22.14-year Hale cycle. 
  
  Inspired by the finding of \cite{Velasco2018}
  that the distribution of Ground Level Enhancement (GLE) events shows phase stability under the assumption of an underlying process with periodicity of 1.73 years, we show in Appendix B that an updated series of GLE events leads to a value of 1.724 years which is even closer to our theoretical value of 1.723 years. Besides the similar issue with the Schwabe cycle, the QBO obviously represents 
  another clocking 
  problem in which the comparison of accurate values is highly relevant.

\begin{figure}[t]
  \centering
  \includegraphics[width=0.99\textwidth]{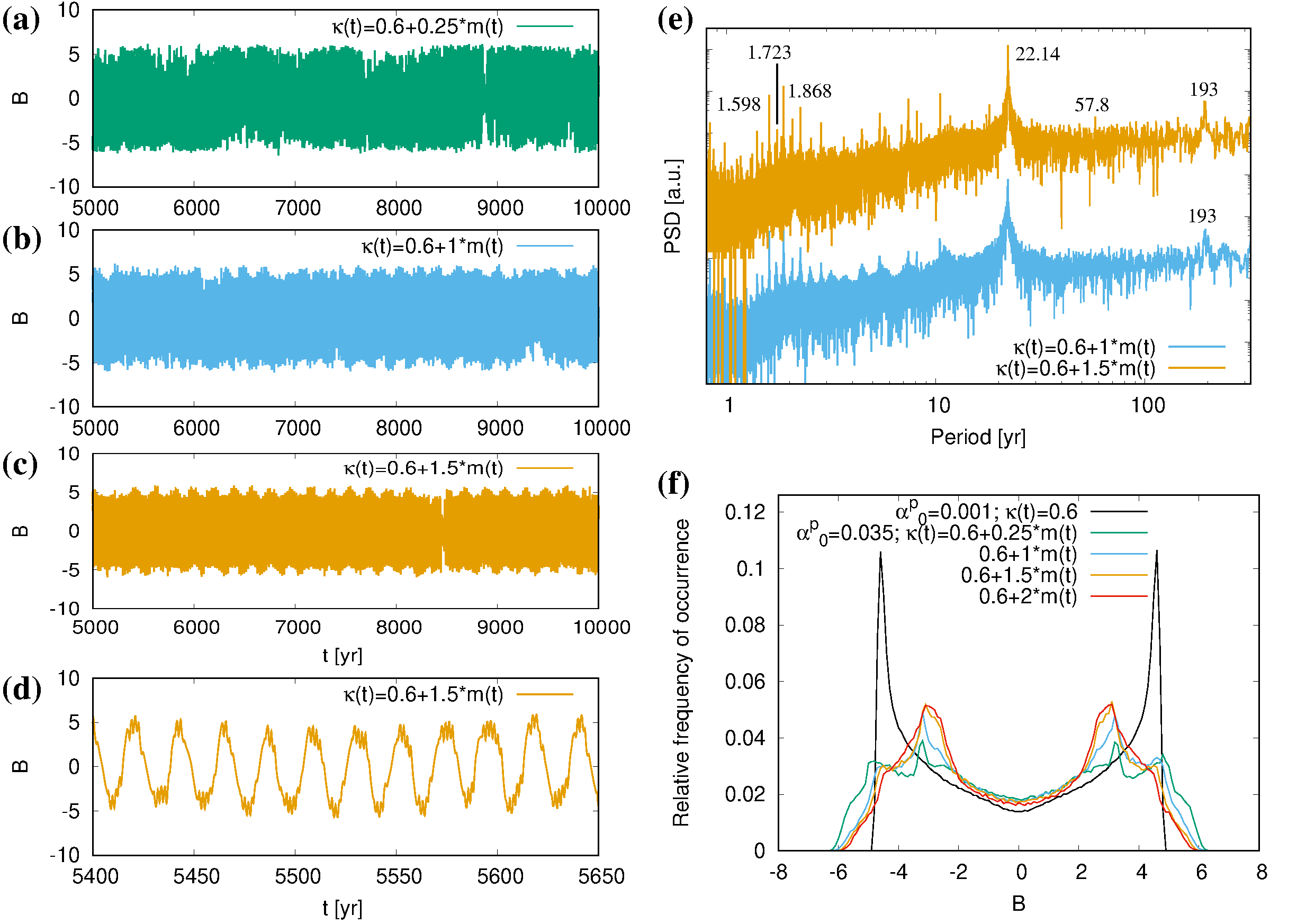}
  \caption{The full model including the QBO, the Hale and the Suess de Vries cycle. (a-c)  Time evolution of $B$ in dependence on the variational factor $\kappa_1$ of the periodic forcing. (d) Detail of (c) over 250 years, showing the Schwabe cycle with the QBO. (e) Spectra of the $B$-fields from (b) and (c), with the typical 
  periods indicated. The type of spectrum, which had been shown in Figure 9 of \cite{Stefani2024} to remarkably agree with climate-related data, is now complemented by the QBO.
  (f) Histograms of $B$ for the three values of $\alpha_0^p$ from (a-c), plus one for $\kappa_1=2$, 
  showing a significant flattening and an increasing second 
  peak at intermediate field strengths.}
\end{figure}

Up to this point we have not included any 19.86-year periodicity, i.e.
$\kappa(t)$ was chosen as constant. We change this now by letting $\kappa$ vary in time.
In Figures 5a-d we show $B$ for increasing strength 
of $\kappa_1$ (additionally we set here the noise-term to zero, $D=0$). The 193-year beat 
period becomes increasingly dominant again. What is remarkable in the 
resulting spectrum in Figure 5e is that {\it all relevant 
periods are present now}, starting from the QBO at
1.723 years (and its side bands), via the Hale cycle to the Suess-de Vries cycle.
Note that the refinement of the model on the QBO time-scale has not much influenced 
the appearance of the long-term periods. Also remarkable are the histograms in 
Figure 5f which show now a drastic smoothing and 
two clearly separated peaks, quite reminiscent of those of \cite{Nagovitsyn2017}. 
Evidently, the dynamo spends
much less time at high magnetic fields. This might well explain the 
subdued character of solar activity as observed by \cite{Reinhold2020}.

We refrain here from elaborating on the possible emergence 
of the even longer cycles of the Bray-Hallstatt type
\citep{Scafetta2016,Scafetta2022}, but turn instead
to a somewhat surprising effect  that might also have some consequences for 
the long term behaviour.

\section{Cycle anomalies and phase jumps}

Anomalies and phase jumps 
are interesting phenomena which have often been discussed 
in connection with the solar dynamo, not least in connection with the
so-called ``lost cycle'' around 1795 \citep{Usoskin2002} (and a similar 
one around 1563, see  \cite{Link1978} and 
\cite{Schove1979}).
Usually, one considers the case that after some anomaly
the dynamo settles back with the same phase as before, or with 
a 180$^{\circ}$ phase shift (the latter would not be visible
in observations based on field intensities alone).
At any rate, it is important to note that the existence of
such phase jumps by no means contradicts
the general concept of synchronization \citep{Pikovsky2003}.

Yet another type of phase jumps was discussed by \cite{Vos2004} in connection with
algae-related data stemming from Lake Holzmaar and Greenland. At five positions 
in the interval 10000-9000 cal. BC, they had observed phase jumps by
90$^{\circ}$, which were attributed to the biological optimality criterion
of the growth of the investigated algae. 
Further below, we will assess an alternative explanation for 
those phase jumps.

Let us start, however, with the first type. Figure 6e shows again the magnetic 
field $B(\theta=72^{\circ},t)$
for a model corresponding to that underlying
Figure 4, but now with
a slightly increased periodic forcing $\alpha^p_0=0.045$ (and 
time-independent $\kappa=0.6$).
Obviously, the emerging time evolution over 2000 years is punctuated by a 
couple of irregularities. Two of those are analyzed in more
detail. In either case the function $h(t)$ is the same as previously,
incorporating the strong 1.723-year QBO-type period and the 11.07-year envelope. In 
the first interval, between the years 24350 and 24650, we observe that
after the anomaly 
the field comes back with the same phase as before (easily visible from the 
comparison with a sine-function with period
22.14 years). In contrast to that the irregularity in the second 
interval, between 25400 ad 25700, leads to a phase shift by 180$^{\circ}$.

\begin{figure}[t]
  \centering
  \includegraphics[width=0.99\textwidth]{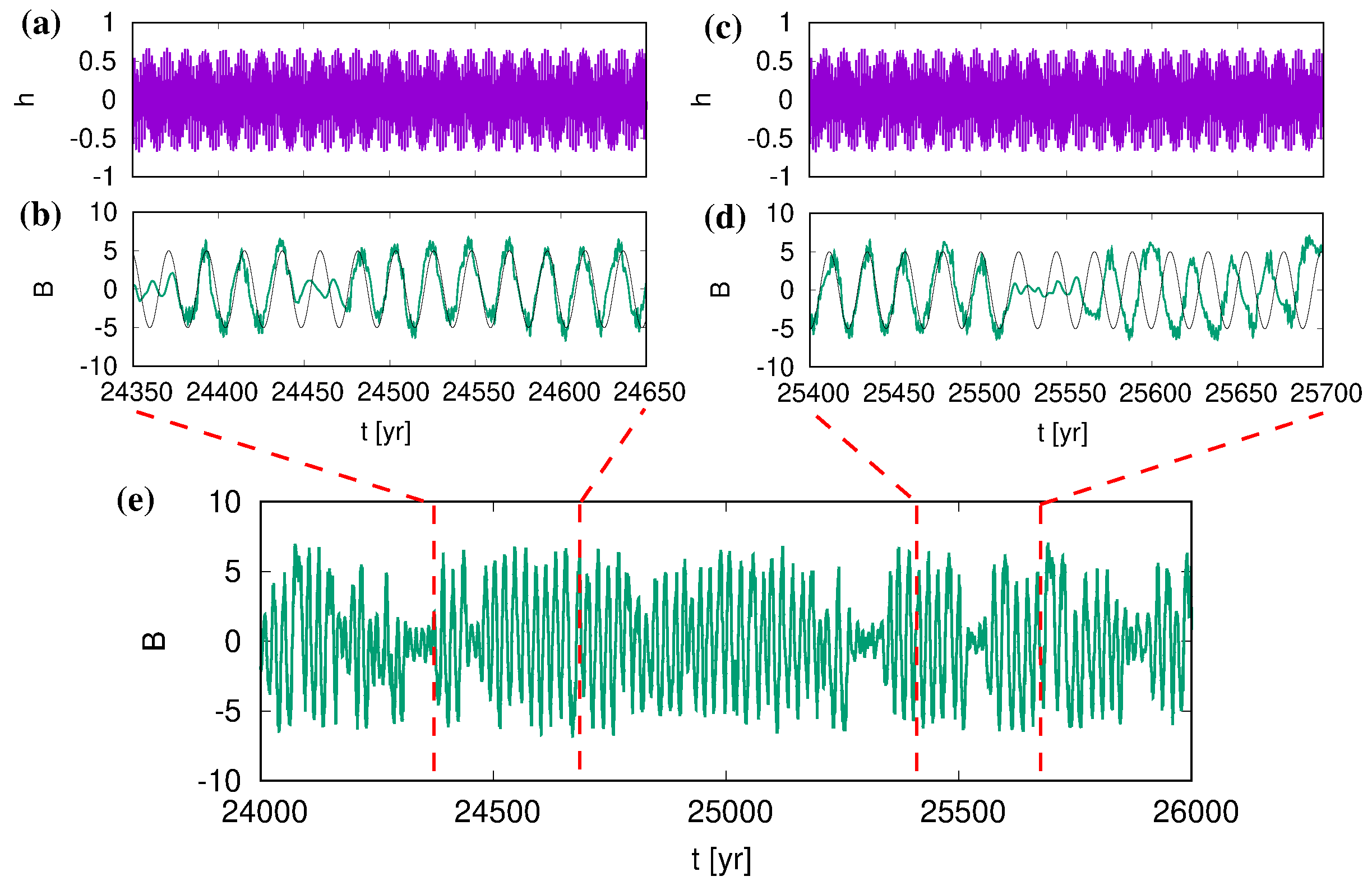}
  \caption{Emergence of irregularities and phase jumps. (a) The function $h(t)$ with 
  the dominant 1.723-year signal and the typical envelope structure with 11.07-year 
  periodicity.
  (b) Field $B(t)$ for the time segment between 24350-24650 years. Note the short breakdown, 
  followed by an evolution with zero phase shift, as evidenced by the overlaid sine-function 
  with 22.14-year periodicity (in gray). (c) and (d) The same as (a) and (b) but for the time segment
  between 25400-25700 years. After the breakdown, we observe a clear 180$^{\circ}$ phase shift.
  (e) Field $B(t)$ for the longer time segment between 24000-26000 years. Note the 
  significant number of irregularities, which are partly connected with phase shifts.}
\end{figure}

Now let us discuss the second possibility of phase jumps by 90$^{\circ}$.
Such phase jumps usually do not occur if the function $h(t)$ is not altered. However, 
as seen in the previous section, the position of the maximum of the 11.07-year envelope 
depends on the width of the moving average window, and can shift by half of that period 
(compare Figure 3d with 3e).

In Figure 7 we demonstrate the possibility that 
such 90$^{\circ}$  phase jumps can indeed occur. 
For that purpose, we have chosen an interval the first part of which (until 30000 years) is 
governed by the function $h(t)$ taken from Figure 3e, while at later times $h(t)$ corresponds 
to that of Figure 3d (in either case with the mean value subtracted beforehand). The other 
parameters are like in Figure 4, 
with $\alpha^p_0=0.03$.

Admittedly, in the first segment the field is not always synchronized and contains a lot 
of irregularities. At least we see a ``successful'' synchronization in the initial 
sub-segment between 29850 and 29920 years. Interestingly, though, after crossing the 
switch between the two $h(t)$ functions (indicated by the red dashed line), the signal 
becomes quickly synchronized again, but now 
with a phase jump of only 90$^{\circ}$.
We are far from claiming that this is indeed the scenario underlying the 90$^{\circ}$ 
phase jumps observed by \cite{Vos2004}, but it might be 
kept in mind as a  principle possibility.

\begin{figure}[t]
  \centering
  \includegraphics[width=0.8\textwidth]{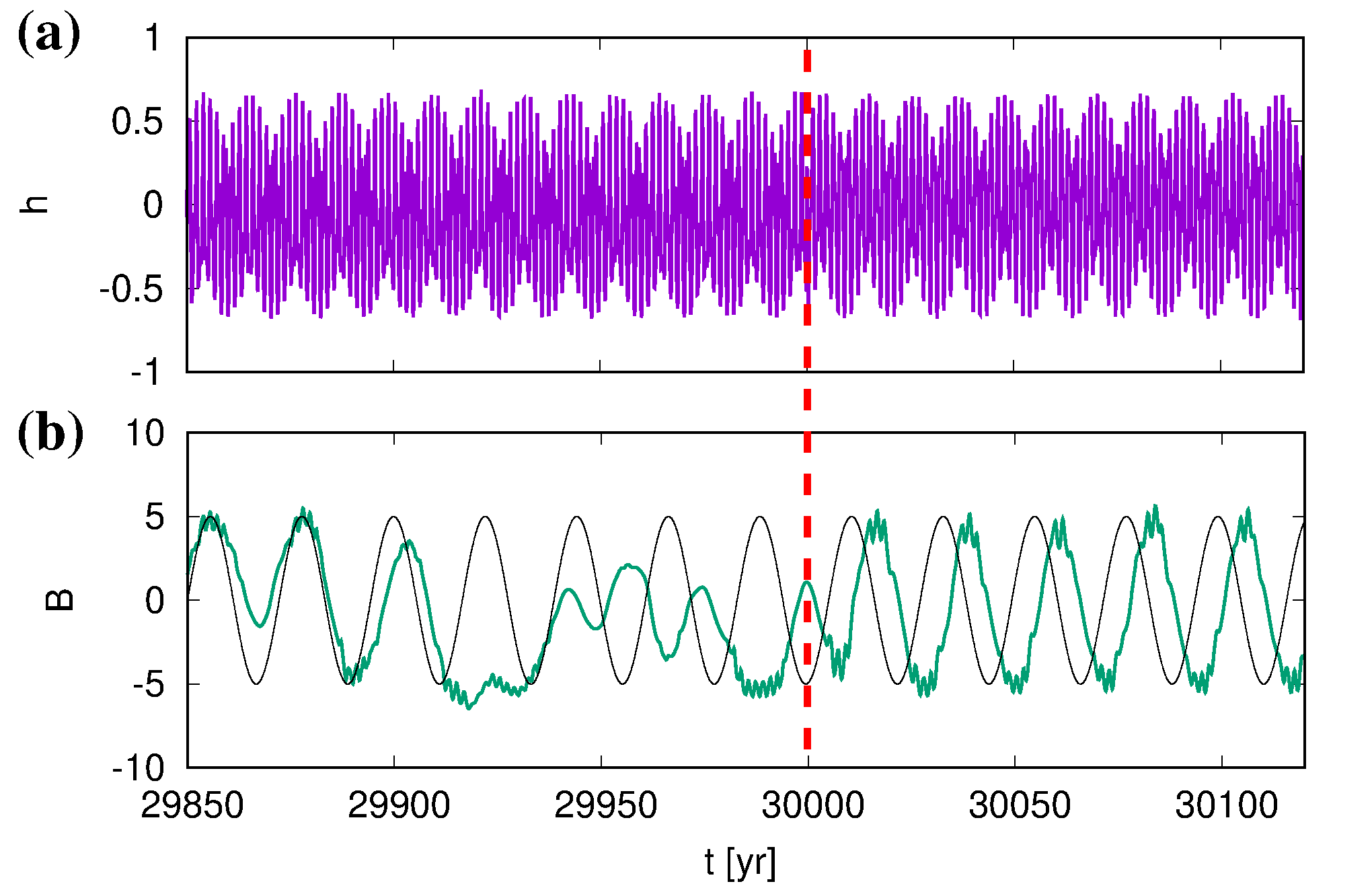}
  \caption{Emergence of a 90$^{\circ}$ phase jump.  (a) The function $h(t)$, comprising one 
  segment corresponding to Figure 3e (before 30000 years), followed by one corresponding to Figure 3d.
  (b) Signal of $B$, with a poor synchronization in the first segment (only in the initial interval), showing a quick resynchronization after the break of year, but now with a 90$^{\circ}$ phase shift. }
\end{figure}

\section{Summary and conclusions}

In this paper, we have continued our efforts to 
corroborate a closed and self-consistent model of 
all observed periods of the solar dynamo in terms 
of synchronization by forces exerted by the orbiting plants.
Building on the previously established link between tidally triggered magneto-Rossby waves 
and Rieger-type periodicities, our main focus 
lied now on the QBO around 1.723 years which is composed of the 199-day and the 292-day 
two-planet spring tide periods. We showed 
this particular QBO to be a key ingredient of any quadratic functional
of the sum of the waves, and noted its remarkable closeness 
to the value of 1.724 years to which we have 
specified
the 1.73-year period 
as previously derived by 
\cite{Velasco2018} from 
Ground Level Enhancement data.

In this framework, the Schwabe period of 11.07 years still emerges as a secondary
beat of the three periods 199, 292, and 625 days. It is the envelope of the superposition 
of these three periods that is still capable of entraining the entire dynamo, by virtue 
of parametric resonance, to a period of 11.07 years, as long as the ``natural'' dynamo 
period is not too far away from
twice that value. The previously found emergence of the Suess-de Vries (and Gleissberg) 
cycle from another beat period of the 22.14-year Hale cycle with 
the 19.86-year periodicity of the Jupiter-Saturn alignments is not seriously 
affected by this model modification.

A remarkable consequences of the emerging QBO is a 
pronounced ``sedation'' of the solar 
dynamo, which now 
spends more time on intermediate field strengths than the undisturbed, single-frequency 
dynamo would do. Indeed, the short-period QBO fosters the longer-term Schwabe cycle 
to ``swing back'' before it reaches its maximum, an effect that also results in a bimodal 
field distribution.  We consider this phenomenon as a promising candidate to explain 
the fact that the solar activity is much more benign than that of other sun-like 
stars \citep{Schaefer2000,Reinhold2020,Cliver2022}. Considering, on one hand, the importance of superflares, 
high UV radiation, high-energy protons, and 
coronal mass ejections for the stability of planetary atmospheres and the development 
of complex live on planets \citep{Karoff2015,Lingam2017}, and on the other hand the 
presumable rarity of the dominant spring-tide periods 
to match the eigenperiods of magneto-Rossby waves, one might even speculate here
about an ``anthropic principle'' of stellar dynamo theory\footnote{In a somewhat bold 
translation of Carter's original definition \citep{Carter1973}, this principle would state 
that ``what type of {\it{planetary system}} we can expect to 
{\it live in} must be restricted by the {\it resulting stellar dynamo} conditions 
necessary for our presence as observers.''
Keeping this ``principle'' in mind may help to avoid possible 
logical fallacies when drawing conclusions for the Sun 
from features observed for other solar-like stars. 
This applies both to the frequency and energy content of superflares
\citep{Vasilyev2024} and to 
the very synchronization of the dominant dynamo cycle 
\citep{Obridko2024}. Since the latter effect additionally requires 
the secondary beat period to be close to half of the typical
period of the unperturbed dynamo, it might occur even less frequently 
than QBO-type cycles which indeed are occasionally 
observed \citep{Olah2009,Soon2019}.}.

In the last part, besides ``normal'' phase jumps by 180$^{\circ}$, we also
discussed the possible emergence of 90$^{\circ}$ phase jumps as a result of the 
remarkable sensitivity of the position of the maxima of the 11-07-year envelope 
on the width of the temporal averaging window. While - as non-biologists - we are not in a position to put into question the plausible argumentation of 
\cite{Vos2004} who had explained those phase jumps in terms of optimum growth conditions of algae, we only cautiously hint at the possibility that an alternative mechanism may exist.

Combining the results of this paper with those of \cite{Stefani2024} (in particular the agreement of the computed spectrum with that derived from climate-related data), we consider the capability of our model to explain  a wide variety of solar dynamo features as 
a solid and promising basis for further advancements. 
While presumably all individual periodicities
could also be explained by choosing appropriate 
sets of dynamo parameters,
in our model they result - in a highly self-consistent 
manner - from the three spring-tide periods 
of Venus, Earth and Jupiter, and the 19.86-year synodic 
period of Jupiter and Saturn. In this sense,
the problem looks to us
like another good candidate for applying Occams's razor.

Having said that, we are well aware of quite a couple of missing pieces in our synchronization  jigsaw, in particular 
the quantitative derivation of various quadratic functionals ($\alpha$, zonal flow, etc.) from the three tidally-excited  magneto-Rossby waves, and 
a deepened understanding of the spin-orbit coupling mechanism underlying the 19.86-year periodicity. Another important point is the confirmation of the results of our simple 1D model by more realistic 2D  or even 3D models
incorporating all nonlinearities in terms of $\alpha$-quenching and 
the $\Lambda$-effect. We cordially invite theoreticians and numericists to join us in this journey.

\begin{acks}
This work received funding from the Helmholtz Association 
in frame of the AI project GEOMAGFOR (ZT-I-PF-5-200), and from 
Deutsche Forschungsgemeinschaft under grant no. MA10950/1-1.
F.S. appreciates financial support by the 
National Science Foundation (Grant No. DMS-1925919), enabling 
a visit at Institute for Pure and Applied Mathematics (IPAM) where
parts of this research were presented.
Many inspiring, if partly controversial, discussions with 
Carlo Albert, Rainer Arlt, J\"urg Beer, Axel Brandenburg, Robert Cameron, 
Christoph Egbers, Chris Jones,
Antonio Ferriz Mas, Uwe Harlander, Bradley Hindman, Fadil Inceoglu, Laur\`ene Jouve, Martins Klevs, Ulrich von Kusserow,
Henri-Claude Nataf, Markus Roth, G\"unther R\"udiger, Nicola Scafetta, 
Jim Shirley, Dmitry Sokoloff, Rodion Stepanov, Andreas Tilgner,
Steve Tobias, Willie Soon, Victor Manuel Velasco Herrera, and Teimuraz Zaqarashvili on various aspects of solar dynamo synchronization are gratefully 
acknowledged.
\end{acks}

\newpage
\appendix 

\section*{Appendix A}
The square of the triad signal $s(t)$, given by Equation\,(8), initially involves 9 periods, which can be shown through the following decomposition of the signal:
\begin{eqnarray}
	s^2(t) &=& \left[\cos\left(2\pi\frac{t-t_{\rm VJ}}{0.5P_{\rm VJ}}\right) + \cos\left(2\pi\frac{t-t_{\rm EJ}}{0.5P_{\rm EJ}}\right) + \cos\left(2\pi\frac{t-t_{\rm VE}}{0.5P_{\rm VE}}\right)\right]^2 \nonumber\\
	&=& \cos\left(2\pi\frac{t-t_{\rm VJ}}{0.5P_{\rm VJ}}\right)^2 + \cos\left(2\pi\frac{t-t_{\rm EJ}}{0.5P_{\rm EJ}}\right)^2 + \cos\left(2\pi\frac{t-t_{\rm VE}}{0.5P_{\rm VE}}\right)^2 \nonumber \\
	&+& 2\cos\left(2\pi\frac{t-t_{\rm VJ}}{0.5P_{\rm VJ}}\right)\cos\left(2\pi\frac{t-t_{\rm EJ}}{0.5P_{\rm EJ}}\right) \nonumber \\&+& 2\cos\left(2\pi\frac{t-t_{\rm EJ}}{0.5P_{\rm EJ}}\right)\cos\left(2\pi\frac{t-t_{\rm VE}}{0.5P_{\rm VE}}\right)\nonumber \\ 
	& +& 2\cos\left(2\pi\frac{t-t_{\rm VJ}}{0.5P_{\rm VJ}}\right)\cos\left(2\pi\frac{t-t_{\rm VE}}{0.5P_{\rm VE}}\right) \nonumber \\
	&=& \cos\left(2\pi\frac{t(P_{\rm VE} + P_{\rm EJ}) - (t_{\rm EJ}P_{\rm VE} + t_{\rm VE}P_{\rm EJ})}{0.5P_{\rm VE}P_{\rm EJ}}\right)\nonumber \\& +& \cos\left(2\pi\frac{t(P_{\rm VE} - P_{\rm EJ}) - (t_{\rm EJ}P_{\rm VE} - t_{\rm VE}P_{\rm EJ})}{0.5P_{\rm VE}P_{\rm EJ}}\right) \nonumber \\
	&+& \cos\left(2\pi\frac{t(P_{\rm VE} + P_{\rm VJ}) - (t_{\rm VJ}P_{\rm VE} + t_{\rm VE}P_{\rm VJ})}{0.5P_{\rm VE}P_{\rm VJ}}\right)\nonumber \\& +& \cos\left(2\pi\frac{t(P_{\rm VE} - P_{\rm VJ}) - (t_{\rm VJ}P_{\rm VE} - t_{\rm VE}P_{\rm VJ})}{0.5P_{\rm VE}P_{\rm VJ}}\right) \nonumber \\
	&+& \cos\left(2\pi\frac{t(P_{\rm EJ} + P_{\rm VJ}) - (t_{\rm VJ}P_{\rm EJ} + t_{\rm EJ}P_{\rm VJ})}{0.5P_{\rm EJ}P_{\rm VJ}}\right)  \nonumber \\& +&\cos\left(2\pi\frac{t(P_{\rm EJ} - P_{\rm VJ}) - (t_{\rm VJ}P_{\rm EJ} - t_{\rm EJ}P_{\rm VJ})}{0.5P_{\rm EJ}P_{\rm VJ}}\right)  \nonumber \\
	&+& \frac{1}{2}\cos\left(2\pi\frac{2(t-t_{\rm VJ})}{0.5P_{\rm VJ}}\right) + \frac{1}{2}\cos\left(2\pi\frac{2(t-t_{\rm EJ})}{0.5P_{\rm EJ}}\right) + \frac{1}{2}\cos\left(2\pi\frac{2(t-t_{\rm VE})}{0.5P_{\rm VE}}\right)\nonumber \\& +& \frac{3}{2} \;.
    \label{Eq:Sig}
\end{eqnarray}
Obviously, the half-periods of all three synodic periods are generated in the squared signal:
\begin{eqnarray}
P_1 = \frac{1}{4}P_{\rm VJ} = 59.25\,{\rm days}, \; P_2 = \frac{1}{4} P_{\rm EJ} = 99.72\,{\rm days}, \; P_3 = \frac{1}{4} P_{\rm VE} = 145.99\,{\rm days} \;.\nonumber
\end{eqnarray}
Six more periods appear from the composition of two periods each:
\begin{eqnarray}
  &P_4 = \frac{0.5P_{\rm VE}P_{\rm EJ}}{P_{\rm VE}+P_{\rm EJ}} = 118.5\,{\rm days} \qquad P_5 = \frac{0.5P_{\rm VE}P_{\rm EJ}}{P_{\rm VE}-P_{\rm EJ}} = 629.29\,{\rm days} \nonumber \\	
  &P_6 = \frac{0.5P_{\rm VE}P_{\rm VJ}}{P_{\rm VE}+P_{\rm VJ}} = 84.29\,{\rm days} \qquad  P_7 = \frac{0.5P_{\rm VE}P_{\rm VJ}}{P_{\rm VE}-P_{\rm VJ}} = 199.43\,{\rm days} \nonumber \\
    &P_8 = \frac{0.5P_{\rm EJ}P_{\rm VJ}}{P_{\rm EJ}+P_{\rm VJ}} = 74.33\,{\rm days} \qquad P_9 = \frac{0.5P_{\rm EJ}P_{\rm VJ}}{P_{\rm EJ}-P_{\rm VJ}} = 291.96\,{\rm days} \nonumber 
\end{eqnarray}
However, these periods are not all independent from the initial synodic periods in $s(t)$. Specifically, we can identify
\begin{eqnarray}
 P_4 = 0.5 P_{\rm VJ}  \qquad P_7 =  0.5 P_{\rm EJ} \qquad P_9 = 0.5 P_{\rm VE} \;.
 \label{eq:periods}
\end{eqnarray}
Up to this point we can conclude that the quadratic action of the three Rossby waves contains the three synodic periods, their half-periods and three composed periods, among them the QBO period $P_5$, without which the 11.07-year period could not emerge as a beat. Using Equation\,(11), $s^2(t)$ can be simplified to
\begin{eqnarray}
	s^2(t) &=& \cos\left(2\pi\frac{t-t_{\rm VJ}}{0.5P_{\rm VJ}}\right) + \cos\left(2\pi\frac{t(P_{\rm VE} - P_{\rm EJ}) - (t_{\rm EJ}P_{\rm VE} - t_{\rm VE}P_{\rm EJ})}{0.5P_{\rm VE}P_{\rm EJ}}\right) \nonumber \\
	&+& \cos\left(2\pi\frac{t(P_{\rm VE} + P_{\rm VJ}) - (t_{\rm VJ}P_{\rm VE} + t_{\rm VE}P_{\rm VJ})}{0.5P_{\rm VE}P_{\rm VJ}}\right) + \cos\left(2\pi\frac{t-t_{\rm EJ}}{0.5P_{\rm EJ}}\right) \nonumber \\
	&+& \cos\left(2\pi\frac{t(P_{\rm EJ} + P_{\rm VJ}) - (t_{\rm VJ}P_{\rm EJ} + t_{\rm EJ}P_{\rm VJ})}{0.5P_{\rm EJ}P_{\rm VJ}}\right) + \cos\left(2\pi\frac{t-t_{\rm VE}}{0.5P_{\rm VE}}\right)  \nonumber \\
	&+& \frac{1}{2}\cos\left(2\pi\frac{2(t-t_{\rm VJ})}{0.5P_{\rm VJ}}\right) + \frac{1}{2}\cos\left(2\pi\frac{2(t-t_{\rm EJ})}{0.5P_{\rm EJ}}\right) + \frac{1}{2}\cos\left(2\pi\frac{2(t-t_{\rm VE})}{0.5P_{\rm VE}}\right) \nonumber \\&+& \frac{3}{2} \;.
\end{eqnarray}
The quadratic signal therefore also contains the linear synodic periods (and their halfs), which comprise the 11.07-year period as a beat period. In fact, all combinations of $0.5P_{\rm VJ}$, $0.5P_{\rm EJ}$ and $0.5P_{\rm VE}$ alone produce a slight Schwabe beat, but not in a trivial manner. The 11.07-years period emerges, in each case, as secondary beats known in music theory as {\it mistuned consonances}.
For the sake of illustration, let us consider the superposition of two harmonic tones with frequencies $f_1$, $f_2$ and phases $\varphi_1$, $\varphi_2$:
\begin{eqnarray}
   w(t) &=& \cos(2\pi (f_1 t - \varphi_1)) + \cos(2\pi (f_2 t - \varphi_2)) \nonumber \\
   &=& 2\cos\left(2\pi \left(\frac{f_1+f_2}{2}t - \frac{\varphi_1+\varphi_2}{2}\right)\right) \nonumber \\
   && \times \cos\left(2\pi \left(\frac{f_1-f_2}{2}t - \frac{\varphi_1-\varphi_2}{2}\right)\right)\label{eq:dyad}
\end{eqnarray}
The primary beat frequency $f_{\rm pb} = f_1 - f_2$, which is twice the frequency of the second modulating cosine, trivially follows from the trigonometric identity (13). The primary beats of $0.5P_{\rm VJ}$, $0.5P_{\rm EJ}$ and $0.5P_{\rm VE}$ are identical to the periods $P_5$, $P_7$ and $P_9$, showing that beat periods of a signal can manifest themselves as ``true'' periods in the squared signal (visible in the Fourier spectrum).  These primary beats alone cannot explain the observed occurrence of the much longer Schwabe period. However, dyads  can also involve {\it secondary beat frequencies}, occurring whenever a multiple of $f_1$ is close to a multiple of $f_2$, i.e., $f_1 = (m/n)f_2 + \delta f$, where $m$ and $n$ are integers  and $\delta f \ll f_1,f_2$. Secondary beat frequencies are then given by
\begin{eqnarray}
    f_{\rm sb} &=& |m f_2 - n f_1| .\label{eq:Fsb}
\end{eqnarray}
The slowest frequency of the dominant, overarching envelope in the dyad signal is determined by the tuple ($m,n$) of co-prime integers minimizing Equation\,(14). In general, the minimizing tuple cannot be determined analytically, making it necessary to search for it numerically. 
Calculating the amplitude of secondary beats is not a simple task, yet, there is an approximate solution for the envelope \citep{Nuno2024}:
\begin{eqnarray}
 {\rm ENV}[w(t)] &=&  2- A(m,n) +  A(m,n)\left|\cos\left(2\pi \left( \frac{f_{\rm sb}}{2}t - \frac{m \varphi_2 - n\varphi_1}{2}\right)\right)\right|. \label{eq:Env} 
\end{eqnarray}
The beat amplitude $A(m,n)$ depends only on the integers $m$ and $n$ and decreases with increasing ($m,n$), so that secondary beats corresponding to high frequency multiples are not visible in the signal. The amplitude functions $A(m,n)$ can be assessed analytically (to be published elsewhere). Here 
we have calculated all the amplitudes presented below numerically.

After this groundwork, we can now calculate the dominating secondary beat of $s(t)$ and $s^2(t)$. First, we show that all three dyads that can be constructed from $0.5P_{\rm VJ}$, $0.5P_{\rm EJ}$ and $0.5P_{\rm VE}$ have the same dominating 11.07 year secondary beat period. Equation (14) becomes minimal for the following period tuples:
\begin{eqnarray}
P_{\rm sb}({\rm VE,EJ}) &=& \frac{3}{0.5P_{\rm VE}} - \frac{2}{0.5P_{\rm EJ}}, \nonumber \\
P_{\rm sb}({\rm VE,VJ}) &=& \frac{5}{0.5P_{\rm VE}} - \frac{2}{0.5P_{\rm VJ}},\nonumber \\
P_{\rm sb}({\rm VJ,EJ}) &=& \frac{3}{0.5P_{\rm VJ}} - \frac{5}{0.5P_{\rm EJ}}. \label{eq:Dyads}
\end{eqnarray}
Remarkably, all three dominant secondary beats yield exactly the 11.07-year period $P_{\rm sb}({\rm VE,EJ}) = P_{\rm sb}({\rm VE,VJ}) = P_{\rm sb}({\rm VJ,EJ}) = P_{\rm VEJ }$. This follows from Scafetta's formula for the period $P_{\rm VEJ }$ of the recurrence pattern of the triple syzygies \citep{Scafetta2022}:
\begin{eqnarray}
P_{\rm VEJ } = \frac{1}{2}\left[\frac{3}{P_{\rm V}} - \frac{5}{P_{\rm E}} + \frac{2}{P_{\rm J}}\right]^{-1},
\end{eqnarray}
where $P_{\rm V} = 224.701$ days, $P_{\rm E} = 365.256$ days, and $P_{\rm J} = 4332.589$ days are the sidereal orbital periods of Venus, Earth, and Jupiter, respectively. As most easily seen in vector notation, all three beat periods add up exactly to the triple syzygies period:
\begin{eqnarray}
  3(1,-1,0) - 2(0,1,-1) &=& (3,-5,2)  \nonumber \\
  5(1,-1,0) - 2(1,0,-1) &=& (3,-5,2) \nonumber \\
  3(1,0,-1) - 5(0,1,-1) &=& (3,-5,2) \; .\nonumber
\end{eqnarray}
The corresponding approximate beat amplitudes are obtained as $A(3,2) \approx 0.382$, $A(5,2) \approx 0.152$ and  $A(3,5) \approx 0.121$ which amounts to $19.1\,\%$,  $7.6\,\%$ and $6\,\%$ of the total signal, respectively. Clearly, the conjunction of the synodic cycle between Venus and Earth and the synodic cycle between Earth and Jupiter produces the largest amplitude, but in all three dyads $P_{\rm VEJ }$ is manifested as by far the most dominant secondary period. The beat period of $s(t)$ can now easily be calculated by splitting the triad of $0.5P_{\rm VJ}$, $0.5P_{\rm EJ}$ and $0.5P_{\rm VE}$ into a sum of three dyads considered in Equation\,(16), finally yielding the envelope
\begin{eqnarray}
 {\rm ENV}[s(t)] &=&  3 - A_{\rm VEJ} + A_{\rm VEJ}\left|\cos\left(\frac{\pi}{P_{\rm VEJ }}t +
 \varphi_{\rm ENV} \right)\right|, \nonumber \\
 {\rm with} \quad A_{\rm VEJ} &=& 0.5[A(3,2) + A(5,2) + A(5,3)]  \approx 0.327 
\end{eqnarray}
and a certain phase $\varphi_{\rm ENV}$. The proportion of the beat amplitude to the total amplitude amounts to almost exactly $11\,\%$. Therefore, the Schwabe beat in $s(t)$ is less distinctive as when considering the dyad $({\rm VE,EJ})$ alone.  However, the situation changes markedly when we consider the quadratic action $s(t)^2$  actually relevant for dynamo action. Again we can compose the squared triad into sums of dyads, yielding the envelope
\begin{eqnarray}
 {\rm ENV}[s^2(t)] &=&  9 - A_{\rm VEJ} + A_{\rm VEJ}\left|\cos\left(\frac{\pi}{P_{\rm VEJ }}t + \varphi_{\rm ENV}\right)\right|, \nonumber \\
 {\rm with} \quad A_{\rm VEJ} &=& 2 - \frac{1}{3}[A(3,2) + A(5,2) + A(5,3)]^2 \approx 1.857 \; ,
\end{eqnarray}
with an amplitude that now accounts for almost $21\,\%$ of the total signal. 

It remains to calculate the beat of the axisymmetric part of $s(t)^2$. The axisymmetric part can be extracted by taking the phase average
\begin{eqnarray}
S(t) &=& \frac{1}{2\pi}\int_0^{2\pi} \left[\cos\left(2\pi\frac{t-t_{\rm VJ}}{0.5P_{\rm VJ}} + 2\varphi\right) + \cos\left(2\pi\frac{t-t_{\rm EJ}}{0.5P_{\rm EJ}} + 2\varphi\right)\right. \nonumber \\
&+& \left.\cos\left(2\pi\frac{t-t_{\rm VE}}{0.5P_{\rm VE}} + 2\varphi\right)\right]^2{\rm d}\varphi \nonumber \\
&=& \cos\left(2\pi\frac{t(P_{\rm VE} - P_{\rm EJ}) - (t_{\rm EJ}P_{\rm VE} - t_{\rm VE}P_{\rm EJ})}{0.5P_{\rm VE}P_{\rm EJ}}\right) \nonumber \\
&+& \cos\left(2\pi\frac{t(P_{\rm VE} - P_{\rm VJ}) - (t_{\rm VJ}P_{\rm VE} - t_{\rm VE}P_{\rm VJ})}{0.5P_{\rm VE}P_{\rm VJ}}\right) \nonumber \\
&+&  \cos\left(2\pi\frac{t(P_{\rm EJ} - P_{\rm VJ}) - (t_{\rm VJ}P_{\rm EJ} - t_{\rm EJ}P_{\rm VJ})}{0.5P_{\rm EJ}P_{\rm VJ}}\right) + \frac{3}{2}  
\nonumber \\
&=& \cos\left(2\pi\frac{t-t_{\rm EJ}}{0.5P_{\rm EJ}}\right) 
+ \cos\left(2\pi\frac{t-t_{\rm VE}}{0.5P_{\rm VE}}\right) \nonumber  \\
&+& \cos\left(2\pi\frac{t(P_{\rm VE} - P_{\rm EJ}) - (t_{\rm EJ}P_{\rm VE} - t_{\rm VE}P_{\rm EJ})}{0.5P_{\rm VE}P_{\rm EJ}}\right) + \frac{3}{2},
\end{eqnarray}
wherein the phase $2\varphi$ reflects the $m=2$-character of the tidally triggered waves. As can be seen from Equation (20), only the periods $P_5$, $P_7=0.5P_{\rm EJ}$ and $P_9=0.5P_{\rm VE}$ remain in the axi-symmetric signal. The calculation of their envelope is analogous to that of the non-averaged signal. We first search the tuples ($m,n$) corresponding to the dominant secondary beat period of the three dyads pair by pair:
\begin{eqnarray}
P_{\rm sb}({\rm VE,EJ}) &=& \frac{3}{0.5P_{\rm VE}} - \frac{2}{0.5P_{\rm EJ}}, \nonumber \\
P_{\rm sb}({\rm EJ,EJ-VE}) &=&  \frac{1}{0.5P_{\rm EJ}} - 3\left(\frac{1}{0.5P_{\rm EJ}}-\frac{1}{0.5P_{\rm VE}}\right),\nonumber \\
P_{\rm sb}({\rm VE,EJ-VE}) &=& \frac{1}{0.5P_{\rm VE}} - 2\left(\frac{1}{0.5P_{\rm EJ}}-\frac{1}{0.5P_{\rm VE}}\right). 
\end{eqnarray}
Quite as before, all beats are identical with the 11.07-year period $P_{\rm sb}({\rm VE,EJ})=P_{\rm sb}({\rm EJ,EJ-VE})=P_{\rm sb}({\rm VE,EJ-VE})= P_{\rm VEJ}$ what can be proven readily by considering the following superpositions of vector syzygies: 
\begin{eqnarray}
  3[(1,-1,0)  - 2(0,1,-1) &=& (3,-5,2) , \nonumber \\
  1(0,1,-1) - 3[(0,1,-1)-(1,-1,0)] &=& 1(0,1,-1) - 3(-1,2,-1)  = (3,-5,2), \nonumber \\
  1(1,-1,0) - 2[(0,1,-1)-(1,-1,0)] &=& 1(1,-1,0) - 2(-1,2,-1) = (3,-5,2). \nonumber
\end{eqnarray}
The tuples ($m,n$) define the three Schwabe amplitudes $A(3,2) \approx 0.382$,
$A(1,3) \approx 0.382$ and $A(1,2) \approx 1$, from which we can again piece together the beat envelope of the full triad signal:
\begin{eqnarray}
 {\rm ENV}[S(t)] &=& 3 + \frac{3}{2} -  A_{\rm VEJ} + A_{\rm VEJ} \left|\cos\left(\frac{\pi}{P_{\rm VEJ }}t + \varphi_{\rm ENV}\right)\right|, \nonumber \\
 {\rm with} \quad A_{\rm VEJ} &=& 0.5[A(3,2) + A(1,3) + A(1,2)] \approx 0.882\label{eq:ENV(S(t))}.
\end{eqnarray}
The Schwabe amplitude of the phase-averaged, axi-symmetric signal $S(t)$ comprises $19.6\,\%$ of the total amplitude.
All derived envelopes were visualized by the green curves in Figure 3.

\section*{Appendix B}

In this Appendix we reconsider the Ground Level 
Enhancement (GLE) events that had been analyzed previously by \cite{Velasco2018}. These sporadic events 
are related to relativistic solar particles
measured at ground level by a network of cosmic ray 
detectors. Figure 3 of \cite{Velasco2018} revealed 
that the considered 56 GLE events occurred preferentially
in the positive phase of an oscillation with a period
of 
1.73\,years\footnote{While in the text, and the caption of 
their Figure 3, a value of 1.7 years was indicated, 
the authors confirmed that the actual value was indeed 1.73\,years \citep{Velasco2025}, 
which is also in agreement with the dominant periods indicated 
in their Table 1.}.
Contrary to the usual assumption that GLE events are random
phenomena, this observation points to an underlying 
clocked process. 

Thus motivated, we have re-analyzed the sequence of the GLE
events. First we 
updated the  56 events of \cite{Velasco2018} 
to the 71 events as provided
by the official database of neutron monitor count rates 
at https://gle.oulu.fi.
Second, in order to accurately 
infer the best fitting period, 
we have replaced the inverse wavelet method of
\cite{Velasco2018} by computing the correlation 
coefficient ${\rm Corr}=1/71 \sum_{i=1}^{71} \cos(2 \pi t_i/P+ \phi_0)$ of the 71 GLE instants $t_i$, using
cosine functions with different periods $P$ and phases
  $\phi_0$. While ${\rm Corr}$ 
  is not exactly Pearson's empirical 
  correlation coefficient,
  it shares with it the main property of lying between 
  -1 and 1, the latter value occurring only for a perfect 
  match between the events and the maxima of the cosine.
  Figure 8 shows the corresponding results. We can see that
  the highest correlation appears for a period 
  of 629.85 days, which corresponds to 1.724 years. 
  This value is remarkably close 
  to the value of 1.723 years
  that we have derived in this paper as the beat of the three 
  tidally triggered magneto-Rossby waves.

\begin{figure}[t]
  \centering
  \includegraphics[width=0.99\textwidth]{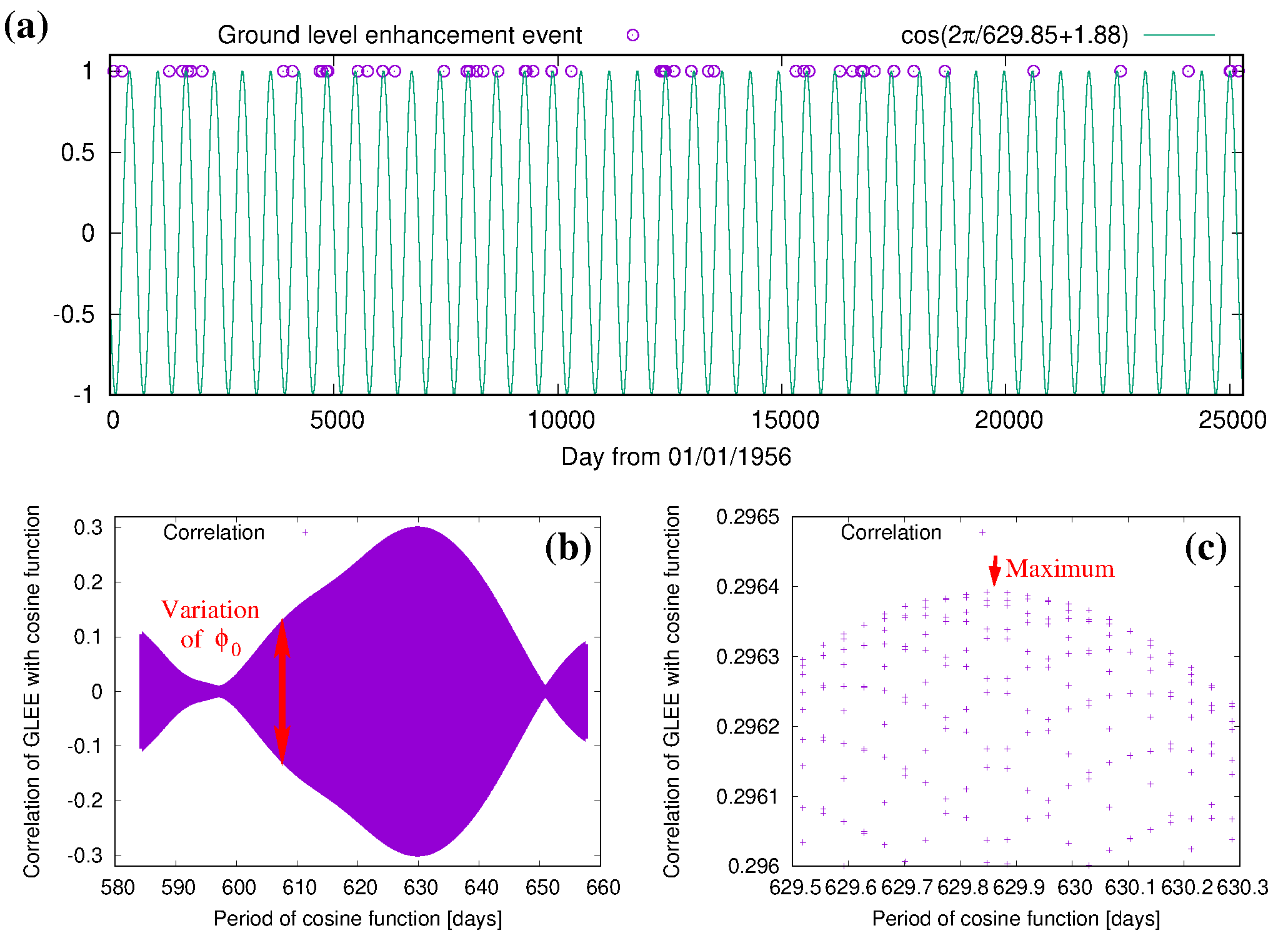}
  \caption{Analysis of Ground Level Enhancement (GLE) events.
  (a) Distribution of 71 GLE events 
  (violet) observed between February 1956 and November 2024, as obtained from https://gle.oulu.fi. 
  The abscissa shows the time $t$ in 
  days from January 1, 1956.
  The green curve shows $\cos(2 \pi t/629.85 + 1.88)$.
  A concentration of the GLE events around the 
  maxima of the cosine function is obvious. 
  (b) Correlation 
  coefficient ${\rm Corr}=1/71 \sum_{i=1}^{71} \cos(2 \pi t_i/P+ \phi_0)$ 
  of cosine functions with different periods $P$ and phases
  $\phi_0$ with the GLE events
  at the 71 instants $t_i$. For each period $P$, the
  vertical extent emerges when using different 
  phases $\phi_0$ between 0 and $2 \pi$.
  (c) Zoomed-in version of (b), showing the maximum 
  of $\rm{Corr}$ at a period of 629.85\,years and for 
  $\phi_0=1.88$. These are the values that are used in the green curve in (a).}
\end{figure}

\section*{Disclosure of Potential Conflicts of Interest}
The authors declare that they have no conflicts of interest.

\end{article} 

\end{document}